\documentclass[10pt,conference]{IEEEtran}
\usepackage{array}
\newcolumntype{P}[1]{>{\centering\arraybackslash}p{#1}}

\usepackage{listings}
\usepackage{xcolor} 
\usepackage{multirow}
\usepackage{booktabs} 
\usepackage{amsmath}
\usepackage{graphicx}
\usepackage{subcaption}
\usepackage{rotating}
\usepackage{tabularx}
\usepackage[noend,linesnumbered, ruled]{algorithm2e}
\usepackage{wrapfig}
\usepackage{listings}
\usepackage{color, colortbl}
\usepackage{balance} 
\usepackage{lscape}
\SetKw{KwStep}{Step}
\SetKw{KwIn}{in}
\usepackage{hyperref}
\usepackage{xspace}
\usepackage{framed}
\usepackage{syntax}
\usepackage{comment}
\usepackage{adjustbox}
\usepackage[flushleft]{threeparttable}
\newcolumntype{C}[1]{>{\centering\arraybackslash}p{#1}}
\usepackage{bbding}
\usepackage{pifont}
\usepackage{wasysym}
\usepackage{tcolorbox}
\usepackage{caption} 
\usepackage{caption} 
\usepackage[framemethod=TikZ]{mdframed}
\usepackage{soul}
\usepackage{flushend}
\usepackage{enumitem}

\usepackage{tikz}

\definecolor{codegreen}{rgb}{0,0.6,0}
\definecolor{codegray}{rgb}{0.5,0.5,0.5}
\definecolor{codepurple}{rgb}{0.58,0,0.82}
\definecolor{backcolour}{rgb}{0.95,0.95,0.92}
\definecolor{Gray}{gray}{0.1}
\lstdefinestyle{mystyle}{
	backgroundcolor=\color{backcolour},   
	commentstyle=\color{codegreen},
	keywordstyle=\color{magenta},
	numberstyle=\tiny\color{codegray},
	stringstyle=\color{codepurple},
	basicstyle=\scriptsize,
	breakatwhitespace=false,         
	breaklines=true,                 
	captionpos=b,                    
	keepspaces=true,                 
	numbers=left,                    
	numbersep=5pt,                  
	showspaces=false,                
	showstringspaces=false,
	showtabs=false,                  
	tabsize=2
}

\lstdefinelanguage{Pythonna}{%
	language     = python,
	morekeywords = {to_categorical, flow_from_directory, pad_sequences, load_image}
}

\lstdefinestyle{customc}{
	belowcaptionskip=1\baselineskip,
	breaklines=false,
	frame= single,
	breaklines = true,
	xleftmargin=\parindent,
	language= Pythonna,
	showstringspaces=false,
	basicstyle=\footnotesize\ttfamily,
	keywordstyle=\bfseries\color{green!40!black},
	commentstyle=\itshape\color{purple!40!black},
	identifierstyle=\color{blue},
	stringstyle=\color{codegreen},
	backgroundcolor=\color{gray!4}
}

\graphicspath{{./figures/}}
\newcommand{\figref}[1]{Fig.~\ref{#1}}

\usepackage{lipsum}
\usepackage[htt]{hyphenat}
\usepackage{float}

\newcounter{rqs}
\stepcounter{rqs}

\newcounter{NumObservations}
\stepcounter{NumObservations}
\definecolor{shadecolor}{rgb}{.9,.9,.9}

\newcommand{\cmark}{\ding{51}\xspace}%
\newcommand{\xmark}{\ding{55}\xspace}%

\AtBeginDocument{%
  \providecommand\BibTeX{{%
    \normalfont B\kern-0.5em{\scshape i\kern-0.25em b}\kern-0.8em\TeX}}}

\begin{document}

\title {Mock Deep Testing: Toward Separate Development of Data and Models for Deep Learning}

\author{\IEEEauthorblockN{
		Ruchira Manke \IEEEauthorrefmark{1},
		Mohammad Wardat \IEEEauthorrefmark{2}, 
		Foutse Khomh \IEEEauthorrefmark{3},
		Hridesh Rajan \IEEEauthorrefmark{1}
	} 
	\IEEEauthorblockA{\IEEEauthorrefmark{1} Dept. of Computer Science, Tulane University, Louisiana, USA, \{rmanke, hrajan\}@tulane.edu}
	\IEEEauthorblockA{\IEEEauthorrefmark{2} Dept. of Computer Science and Engineering,  Oakland University, Michigan, USA, wardat@oakland.edu}
 	\IEEEauthorblockA{\IEEEauthorrefmark{3} SWAT Lab., Polytechnique Montréal, Montréal, Canada, foutse.khomh@polymtl.ca}  	
}

\maketitle
\thispagestyle{plain}
\pagestyle{plain}

\begin{abstract}
While deep learning (DL) has permeated, and become an integral component of many critical software systems, today software engineering research hasn't explored how to separately test data and models that are integral for DL approaches to work effectively. 
The main challenge in independently testing these components arises from the tight dependency between data and models.
This research explores this gap, introducing our methodology of {\em mock deep testing}
for unit testing of DL applications. 
To enable unit testing, we introduce a design paradigm that decomposes the workflow into distinct, manageable components, minimizes sequential dependencies, and modularizes key stages of the DL, including data preparation and model design. For unit testing these components, we propose modeling their dependencies using mocks.
In the context of DL, mocks refer to mock data and mock model that mimic the behavior of the original data and model, respectively.
This modular approach facilitates independent development and testing of the components, ensuring comprehensive quality assurance throughout the development process.
We have developed {\em KUnit}, a framework for enabling mock deep testing
for the Keras library, a popular library for developing DL applications. We empirically evaluated {\em {KUnit}} to determine the effectiveness of mocks in independently testing data and models. Our assessment of 50 DL programs obtained from \sof and \textit{GitHub} shows that mocks effectively identified 10 issues in the data preparation stage 
and 53 issues in the model design stage.
We also conducted a user study with 36 participants using {\em KUnit} to perceive the effectiveness of our approach.
Participants using {\em KUnit} successfully resolved 25 issues in the data preparation stage
and 38 issues in the model design stage.
Our findings highlight that mock objects provide a lightweight emulation of the dependencies for unit testing, facilitating early bug detection.
Lastly, to evaluate the
usability of {\em KUnit}, we conducted a post-study survey. The results reveal that {\em KUnit} is helpful to DL application developers, enabling them to independently test each component (data and model) and resolve issues effectively in different stages.

\end{abstract}


\section{Introduction}
\label{sec:intro}

\textit{Deep Learning} (DL) is a sub-class of machine learning algorithms that has gained a lot of attention from the industry and academia due to its successful adoption in many domains~\cite{Csaky19,Iannizzotto18,Roy18}.
The popularity of DL applications has drawn the interest of the software engineering community and the community has responded by conducting several studies~\cite{zhang2019,Zhang2020,islam19,humbatova20taxonomy,zhang18,cao2022understanding,rahman2023machine} to understand the development process of these applications.
These studies found that DL application developers usually focus on building and optimizing models using the training data, focusing less on modern software engineering practices such as modular design, unit testing, \etc.~\cite{Zhang2020, pan2020decomposing, pan2022decomposing}.
DL application development follows a workflow that is different from the traditional software development~\cite{zhang2019, biswas22art, Amershi2019} - where data is prepared first followed by model design and training, establishing a tight dependency between data and model.
Therefore, incorporating software engineering practices, such as independent testing of data and models necessitates decomposing the workflow, \ie, separating the data and model, and mimicking their dependencies to facilitate unit testing.

Inspired by the fundamental practice of unit testing in traditional software development~\cite{binder2000testing}, and the notion of creating mock objects that mimic the minimum expected behavior of dependencies for unit testing~\cite{mackinnon2000endotesting}, we ask: 
\textit{can we apply the concept of mock objects, commonly used in unit testing traditional software, to test DL applications?}
Unit testing with mock objects not only allows for early bug detection in the development cycle but also facilitates the development of modules and verifying their functionality by deferring the dependencies.
To the best of our knowledge, unit testing using mocks for DL applications — wherein the data and DL model are tested independently — has not been explored before.

This paper introduces the idea of mock testing in the context of \textit{Deep Neural Networks (DNNs)}.
In current DL application development, data is typically prepared by data scientists, while models are designed by machine learning engineers~\cite{Zhang2020}. Each group focuses on distinct stages of the DL pipeline, creating a natural separation of responsibilities. To align with this practice, we recommend treating the data preparation and model design stages as independent modules or units
and propose employing mocks for their independent testing.
In the context of DL, mocks refer
to mock data and mock model that mimic the behavior of the
original data and model, respectively.
The independent testing of these modules using mocks
simplifies debugging, facilitates early bug detection, and ensures that the resulting code meets
certain quality aspects, \ie, good-quality data, a model that conforms with the requirements, and high operational reliability. 

\begin{figure*}[htbp]
	\centering
	\includegraphics[scale=0.6]
 {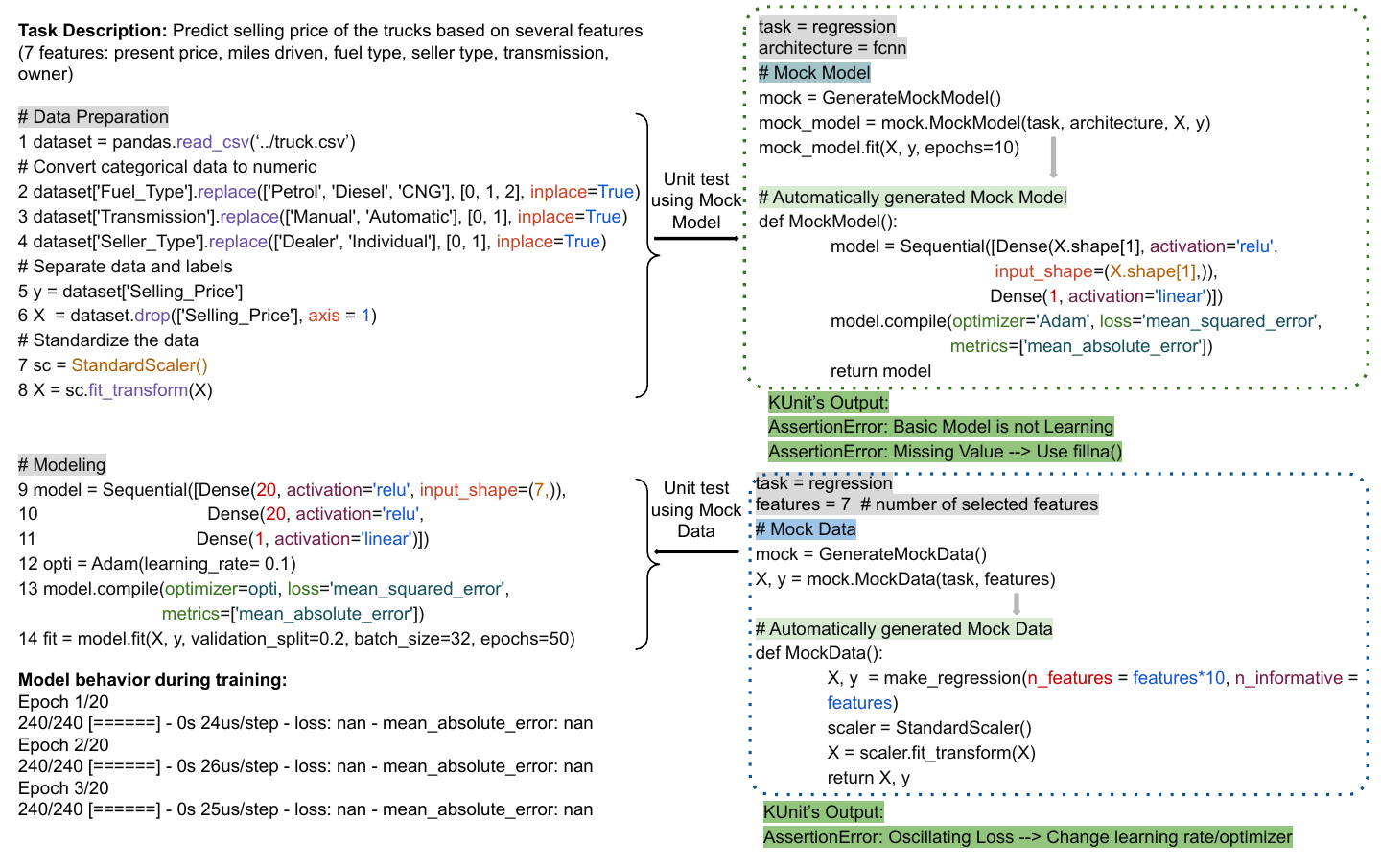}
    \caption{A buggy DL program and mocks in action.}
   
	\label{Motivation}
 \vspace{-0.7cm}
\end{figure*}

To introduce our notion of mock testing, we have focused on two types of DL architectures, Fully-Connected Neural Networks (FCNNs) and Convolution Neural Networks (CNNs) for regression and classification problems.
These architectures are commonly employed to handle high-dimensional data due to their ability to capture nonlinear relationships within datasets~\cite{goodfellow2016deep}.
To facilitate unit testing, we introduce a design paradigm that considers each stage of the DL program, \ie, data preparation and model design, as separate modules. The unique challenge in independently testing these components arises from the tight dependency between data and models. To handle the inherent dependencies among these modules, we propose defining clear interfaces to decouple them. These interfaces specify key elements of each stage, such as the ‘number of features’ in data preparation and the ‘DNN architecture type’ in model design, which influence each other’s configuration. These interfaces are then leveraged to automatically create mock data or models that replicate the behavior of real components. This proposed approach allows for isolated testing of each module by substituting the original data or model with the automatically generated mock versions, ensuring independent quality assurance at each stage. 
To achieve this, we utilized Python's built-in unit testing framework, \texttt{unittest}, and 
developed, \textit{KUnit} specifically for Keras. 
\textit{KUnit} comprises 15 distinct methods with assertions aimed at verifying the expected behavior of specific sections of the code under test, leveraging the generated mocks.
\textit{KUnit} is open-sourced~\cite{myRepo} and can be extended to incorporate more assertions and support other frameworks.

\begin{figure*}[!ht]
	\centering
	\includegraphics[scale=0.65]
 {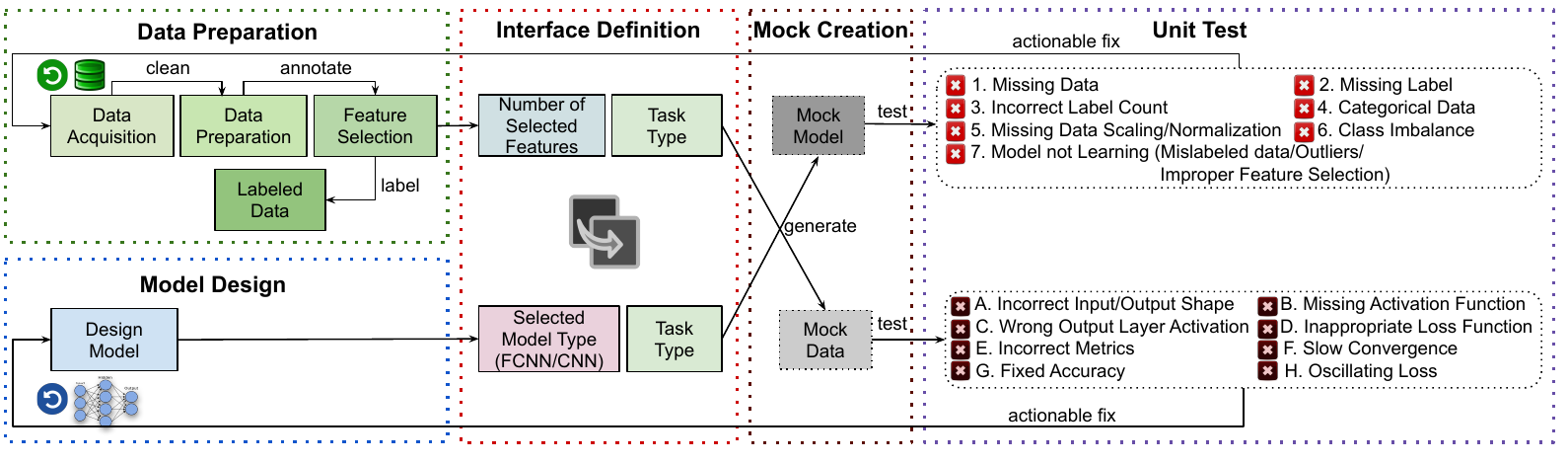}
    \caption{Workflow of {\em KUnit}. }
   
	\label{Overview}
 \vspace{-0.55cm}
\end{figure*}

We have evaluated {\em {KUnit}} through empirical and user evaluation. Empirical evaluation is performed on 50 programs obtained from \sof and \textit{GitHub}.
We separated the data preparation and model design
steps into two distinct modules, which were then tested in isolation using mocks.   
We compared the issues detected using mocks with issues detected when analyzing the two stages together.
We observed that, for the data preparation stage, the mock model helped identify 10 issues, whereas, for the model design stage, the mock data assisted in identifying 53 issues during the empirical evaluation. 
Our results demonstrate that mocks effectively detect issues that cause abnormal behavior during training.
We also performed a user study with 36 participants using {\em KUnit} to evaluate its effectiveness. 
Participants using {\em KUnit} successfully resolved 25 issues in the data preparation stage and 38 issues in the model design stage.
Our findings indicate that mock objects provide an effective, lightweight simulation of dependencies for unit testing. 
In the post-study survey, we found that {\em KUnit} is helpful to the developers for testing each component independently and resolving issues early in the development process.

In summary, our work makes the following 
contributions:
\begin{enumerate}
\item \textbf{Originality:}
First, we introduce mock deep testing for independent testing of data and model.
Next, we identify the elements of each stage that are used to decouple the data preparation and model design stages. Leveraging these elements defined in an interface, we propose a method for automatically generating mock objects for each stage. These mock objects support unit testing and help identify issues early in the development process.

\item \textbf{Usefulness:}
We develop a framework, {\em{KUnit}}, that is extensible and generalized to different classes of DL bugs.
We specified
15 different bug types and the conditions necessary for their detection. 
These conditions are incorporated as assertions in the test methods to identify various bugs and repair strategies are proposed 
as actionable fixes in {\em{KUnit}}.

\item \textbf{Evaluation:}
The empirical and user evaluation shows mock objects' potential for unit testing DL applications. Our results show that mock objects provided a lightweight emulation of dependencies, allowing early bug detection.
The user evaluation provides evidence that mocks are helpful to developers in testing each component independently and resolving issues effectively.

\end{enumerate}

\section{Motivation}
\label{sec:background}

The current practice in DL application development involves sequentiality, where the data is prepared first followed by model design and training. 
The designed model is tested for 
crash and silent bugs using the data by monitoring and identifying abnormal behavior during training~\cite{schoop2021umlaut,wardat21DeepLocalize,wardat22DeepDiagnosis,BraiekDeepChecker,cao2022deepfd,Zhang21Autotrainer,ghanbari2023deepmufl}.
Bugs can originate from any stage of the DL pipeline, such as data preparation or model design, and often exhibit similar symptoms during training.
For example, exploding gradients, a common issue that can arise from the data preparation stage due to improper training data or from the model design stage due to high learning rate, improper weight or bias initialization, and large batch size~\cite{wardat22DeepDiagnosis}. This overlap in the symptoms makes it challenging to pinpoint the root cause of the bugs~\cite{islam20repairing}, thereby requiring several iterations to identify the stage of origin of the bug correctly
\cite{schoop2021umlaut,wardat21DeepLocalize,wardat22DeepDiagnosis,BraiekDeepChecker,ghanbari2023deepmufl}.
This highlights the need for a systematic approach for testing and debugging each stage in isolation.

In traditional software development, unit testing has proven useful for conducting a lightweight evaluation of each functionality in isolation. It holds the potential to identify areas for improvement before integrating the two functionalities~\cite{runeson2006survey}.
Due to the dependency of the model design stage on the data preparation stage in the DL applications, unit testing cannot be applied directly.
Our insight is that by decoupling these stages and using the concept of mocks, each module can be tested independently before integration.
In the context of DL, \textit{mock data} refers to synthetic data designed to replicate the key characteristics of real data, while a \textit{mock model} is a simplified version of the real model that mimics its behavior without incorporating the complexities of the full model architecture.

To illustrate, consider a DL program shown in \figref{Motivation} designed to predict the selling price of the trucks based on 7 features. 
The code from \texttt{Lines 1-8} represents the preparation of the data, \texttt{Lines 9-13} represents the model design, and finally the model is trained (\texttt{Line 14}) using the data obtained from \texttt{Lines 5 \& 8}.
During model training, the program behaved erratically, resulting in NaN values for both the metrics, \ie, mean\_squared\_error and mean\_absolute\_error.
NaN loss during training can arise from either of the two stages: the data preparation stage, due to NaN values in data, or the model design stage, due to too high learning rate causing model parameters to update too aggressively, divide by zero error during learning or incorrect weight initialization~\cite{wardat21DeepLocalize}. 
For instance, this behavior occurred because of the NaN value in the data for the DL program shown in \figref{Motivation}, as the developer forgot to remove or replace missing values during data preparation.
Even if the issues in the data preparation stage are addressed, silent bugs in the model design stage can still occur.
Isolating the two stages and testing them independently using mocks can help the developer identify and address the issue at the correct stage,
thereby reducing the overall debugging effort required during the training process.
This motivates the development of {\em KUnit}, a novel approach for facilitating unit testing of DL applications using mocks.
The fundamental idea of {\em KUnit} is based on the observation that the behavior of the original data on the mock model and the original model on the mock data remains consistent.
To illustrate, for the example in \figref{Motivation}, the unit testing of the data preparation stage using a mock model resulted in NaN loss
(consistent with the original model behavior). 
{\em KUnit} reported that the issue occurred because of missing values in the data which can be addressed in the data preparation stage.
Similarly, for the model design stage, the oscillating loss on the mock data reported by {\em KUnit} (consistent with the model behavior on original data after replacing the missing values), indicates incorrect hyperparameter selection, which can be refined before combining two stages.
Unit testing these stages and resolving issues at the error-inducing stage enables early resolution of potential problems before initiating the training process.
The rest of this work describes our approach, {\em KUnit}, for enabling mock testing for DL applications.

\begin{figure*}[t!]
    \centering
    \subfloat[\centering Interface of data preparation stage]{{\includegraphics[height=4.5cm]{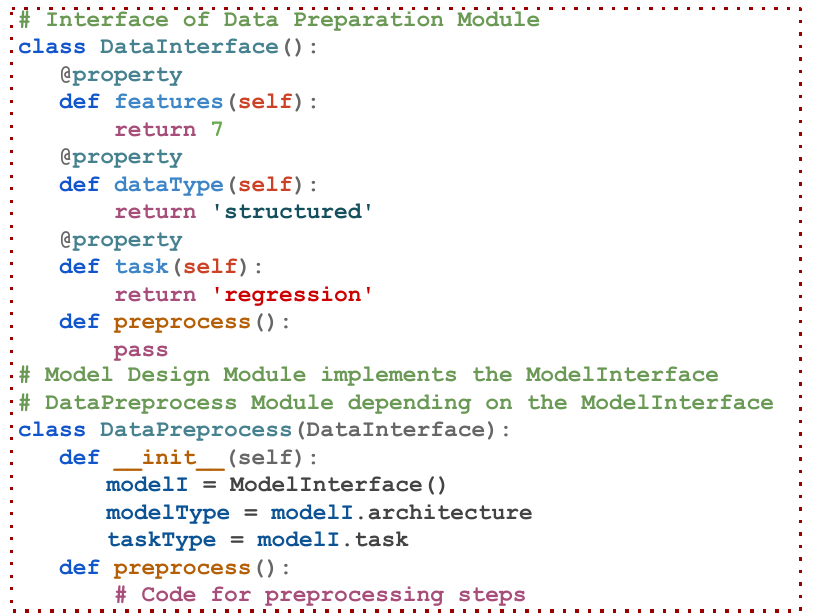} }}%
    \qquad
    \subfloat[\centering Interface of model design stage]{{\includegraphics[height=4.5cm]{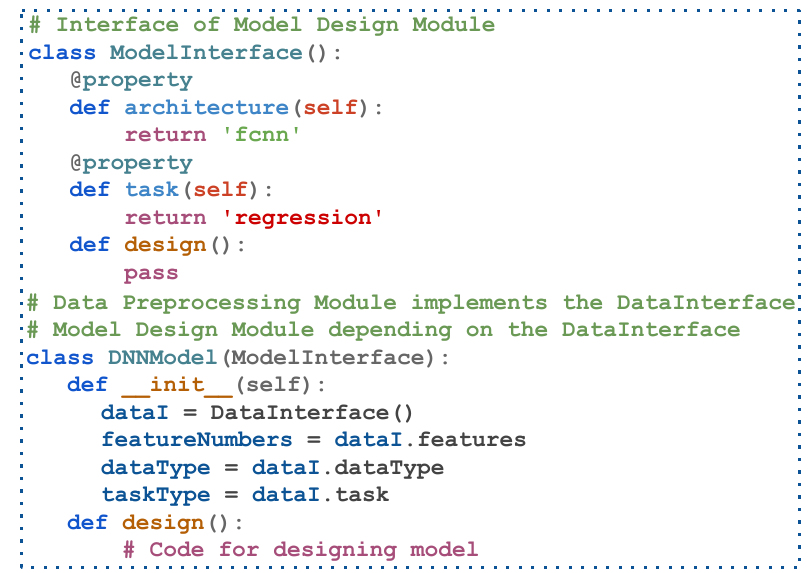} }}%
    \caption{Interface definition and class description.}%
    \label{fig:InterfaceDefinition}
\vspace{-0.6cm}
\end{figure*}

\section{Approach}
\label{sec:Approach}

In this section, we provide an overview of our approach for unit testing DL applications using mocks.
Inspired by the decomposition criteria proposed by Parnas~\cite{parnas1972criteria}, we suggest making each major step in the DL program a module.
In a DL pipeline, these major steps correspond to the different stages, \ie, \textit{data preparation} and \textit{model design}.
Due to coupling between these stages, we decouple them by defining interfaces, allowing the data preparation and model design stages to depend on the interface, ensuring their independence. 
These interfaces facilitate the automatic generation of mocks, which are used for unit testing of each stage.
The workflow of our approach, {\em KUnit} is shown in \figref{Overview}.
We collected issues for each stage from various sources outlined in Section~\ref{bugconditions} and established the conditions for identifying them.
In total, we obtained 7 issues (1-7) for the data preparation
and 8 issues (A-H) for the model design stage shown in \figref{Overview}. 
We 
leveraged Python’s built-in unit testing framework, \texttt{unittest},
to develop, {\em KUnit}; a testing framework for Keras.
We defined each condition as an assertion in the test method aimed at verifying the expected behavior of the code under test leveraging the generated mocks.
Once a failure is detected by {\em KUnit}, the user is notified with an assertion error and a workable solution. 
Our approach 
for unit testing DL applications has two main steps: \textit{interface definition} and \textit{mock object creation and verification}. 
Below, we discuss each step in detail.

\subsection{Interface Definition}
Due to dependencies between the data preparation and model design stages, the primary task for independent testing of these stages is to design interfaces that decouple them. 
For decoupling, it's essential to identify elements from one stage that affect the design decisions of the other.
Understanding these dependencies enables better modularization and smoother integration between stages.
We propose interfaces that allow data preparation and model design stages to depend on the interface, ensuring their independence.
Below, we detail our approach to defining interfaces for each stage.

\paragraph{Interface for Data Preparation Stage}
In the data preparation stage, feature engineering is a common activity carried out intending to select informative features that the DL model learns during training.
These features play a crucial role in the design decisions of the model design stage.
For example, in the model design stage, most of the decisions, such as which neural network architecture to choose and its hyperparameters depend on the characteristics of data and the features selected in the data preparation stage. 
For instance, consider a model in \figref{Motivation} designed for predicting the selling price of trucks.
As the dataset is structured (each row representing a different record), 
the developer selected the FCNN model, which is known to perform well for structured data~\cite{borisov2022deep}.
Since this is a regression task, choosing the appropriate hyperparameters is another design decision in the model design stage.
For example, the output layer activation function depends on the type of task, \ie, regression or classification.
For the data preparation stage, we propose an interface that incorporates the number of features selected during the feature selection step, the type of data, and the type of task.
For the task in \figref{Motivation}, \figref{fig:InterfaceDefinition}(a) shows the interface and class description of the data preparation stage. 
The data preparation module implements this interface, exposing its behavior to other classes that depend on it for their design decisions.

\begin{figure*}[t!]
    \centering
    \subfloat[\centering Mock model creation for data preparation stage]{{\includegraphics[height=3.5cm]{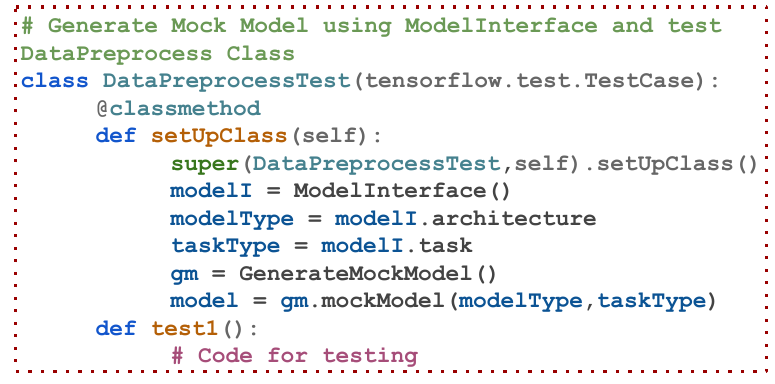} }}%
    \qquad
    \subfloat[\centering Mock data creation for model design stage]{{\includegraphics[height=3.5cm]{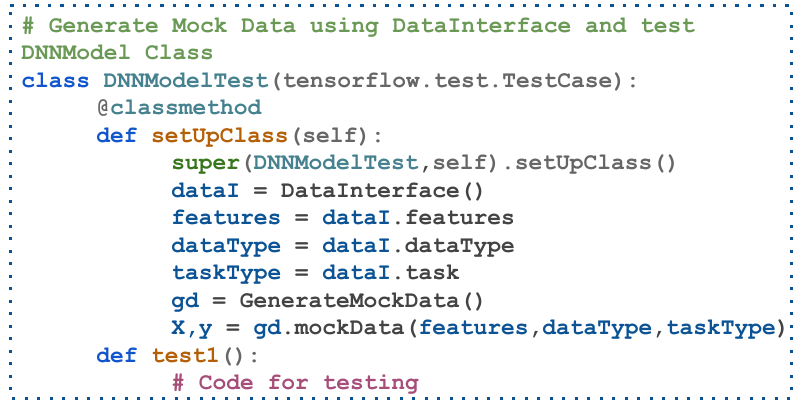} }}%
    \caption{Mock object creation for different stages.}%
    \label{fig:MockCreation}
\vspace{-0.5cm}
\end{figure*}

\paragraph{Interface for Model Design Stage}
In the model design stage, selecting an appropriate neural network architecture corresponding to the task is crucial for achieving optimal performance. 
For instance, FCNNs are a good choice for tasks involving structured data~\cite{borisov2022deep}, where each feature is independent and there are no inherent spatial or temporal relationships to consider, whereas CNNs are well-suited for image classification tasks due to their ability to capture spatial hierarchies in images~\cite{lecun2015deep}.
In the data preparation stage, feature selection is influenced by the neural network architecture chosen in the model design stage. 
Since different architectures have different learning processes, the informative features 
are usually refined based on the model's performance during evaluation~\cite{biswas22art}.
For instance, in FCNNs there is no inherent weight sharing 
whereas, CNNs have a key feature of weight sharing through the convolution filters.
As a result, these two architectures may perform differently even with the same data. 
Therefore, for the model design stage, we propose an interface comprising the architecture chosen in this stage and the type of task.
\figref{fig:InterfaceDefinition}(b) shows the interface and class description of the model design stage for the task in \figref{Motivation}.
The model design module implements this interface and its behavior is exposed to other classes that depend on it.
The defined interfaces are utilized for creating useful mocks and testing each module independently.

\begin{table}[]
\renewcommand{\arraystretch}{1.1}
\caption{{\em KUnit}'s mock model generation process.}
 \centering
 \scalebox{0.48}{
\begin{adjustbox}{width=1.0\textwidth}


\begin{tabular}{|ccl|cccccc|}
\hline
\multicolumn{3}{|c|}{\textbf{}}                                                                                                                         & \multicolumn{6}{c|}{\textbf{Rules}}                                                                                                                                                                                                                                                                                                                                                                                                                                                                                                                                        \\ \hline
\multicolumn{1}{|c|}{}                                      & \multicolumn{1}{c|}{}                                  & Regression                       & \multicolumn{1}{c|}{\cellcolor[HTML]{B5E6A2}Yes}      & \multicolumn{1}{c|}{\cellcolor[HTML]{B5E6A2}Yes}      & \multicolumn{1}{c|}{\cellcolor[HTML]{FBE2D5}No}                                                                 & \multicolumn{1}{c|}{\cellcolor[HTML]{FBE2D5}No}                                                                 & \multicolumn{1}{c|}{\cellcolor[HTML]{FBE2D5}No}                                                                      & \cellcolor[HTML]{FBE2D5}No                                                                      \\ \cline{3-9} 
\multicolumn{1}{|c|}{}                                      & \multicolumn{1}{c|}{}                                  & Binary Classification            & \multicolumn{1}{c|}{\cellcolor[HTML]{FBE2D5}No}       & \multicolumn{1}{c|}{\cellcolor[HTML]{FBE2D5}No}       & \multicolumn{1}{c|}{\cellcolor[HTML]{B5E6A2}Yes}                                                                & \multicolumn{1}{c|}{\cellcolor[HTML]{B5E6A2}Yes}                                                                & \multicolumn{1}{c|}{\cellcolor[HTML]{FBE2D5}No}                                                                      & \cellcolor[HTML]{FBE2D5}No                                                                      \\ \cline{3-9} 
\multicolumn{1}{|c|}{}                                      & \multicolumn{1}{c|}{\multirow{-3}{*}{Problem Type}}    & Multiclass Classification        & \multicolumn{1}{c|}{\cellcolor[HTML]{FBE2D5}No}       & \multicolumn{1}{c|}{\cellcolor[HTML]{FBE2D5}No}       & \multicolumn{1}{c|}{\cellcolor[HTML]{FBE2D5}No}                                                                 & \multicolumn{1}{c|}{\cellcolor[HTML]{FBE2D5}No}                                                                 & \multicolumn{1}{c|}{\cellcolor[HTML]{B5E6A2}Yes}                                                                     & \cellcolor[HTML]{B5E6A2}Yes                                                                     \\ \cline{2-9} 
\multicolumn{1}{|c|}{}                                      & \multicolumn{1}{c|}{}                                  & FCNN                             & \multicolumn{1}{c|}{\cellcolor[HTML]{B5E6A2}Yes}      & \multicolumn{1}{c|}{\cellcolor[HTML]{FBE2D5}No}       & \multicolumn{1}{c|}{\cellcolor[HTML]{B5E6A2}Yes}                                                                & \multicolumn{1}{c|}{\cellcolor[HTML]{FBE2D5}No}                                                                 & \multicolumn{1}{c|}{\cellcolor[HTML]{B5E6A2}Yes}                                                                     & \cellcolor[HTML]{FBE2D5}No                                                                      \\ \cline{3-9} 
\multicolumn{1}{|c|}{}                                      & \multicolumn{1}{c|}{\multirow{-2}{*}{Model Type}}      & CNN                              & \multicolumn{1}{c|}{\cellcolor[HTML]{FBE2D5}No}       & \multicolumn{1}{c|}{\cellcolor[HTML]{B5E6A2}Yes}      & \multicolumn{1}{c|}{\cellcolor[HTML]{FBE2D5}No}                                                                 & \multicolumn{1}{c|}{\cellcolor[HTML]{B5E6A2}Yes}                                                                & \multicolumn{1}{c|}{\cellcolor[HTML]{FBE2D5}No}                                                                      & \cellcolor[HTML]{B5E6A2}Yes                                                                     \\ \cline{2-9} 
\multicolumn{1}{|c|}{}                                      & \multicolumn{1}{c|}{}                                  & Number of classes = 1            & \multicolumn{1}{c|}{\cellcolor[HTML]{B5E6A2}Yes}      & \multicolumn{1}{c|}{\cellcolor[HTML]{B5E6A2}Yes}      & \multicolumn{1}{c|}{\cellcolor[HTML]{FBE2D5}No}                                                                 & \multicolumn{1}{c|}{\cellcolor[HTML]{FBE2D5}No}                                                                 & \multicolumn{1}{c|}{\cellcolor[HTML]{FBE2D5}No}                                                                      & \cellcolor[HTML]{FBE2D5}No                                                                      \\ \cline{3-9} 
\multicolumn{1}{|c|}{}                                      & \multicolumn{1}{c|}{}                                  & Number of classes = 2            & \multicolumn{1}{c|}{\cellcolor[HTML]{FBE2D5}No}       & \multicolumn{1}{c|}{\cellcolor[HTML]{FBE2D5}No}       & \multicolumn{1}{c|}{\cellcolor[HTML]{B5E6A2}Yes}                                                                & \multicolumn{1}{c|}{\cellcolor[HTML]{B5E6A2}Yes}                                                                & \multicolumn{1}{c|}{\cellcolor[HTML]{FBE2D5}No}                                                                      & \cellcolor[HTML]{FBE2D5}No                                                                      \\ \cline{3-9} 
\multicolumn{1}{|c|}{\multirow{-8}{*}{\multirow{-2}{*}{\rotatebox{90}{\textbf{Conditions}}}}} & \multicolumn{1}{c|}{\multirow{-3}{*}{Classes}}         & Number of classes \textgreater 2 & \multicolumn{1}{c|}{\cellcolor[HTML]{FBE2D5}No}       & \multicolumn{1}{c|}{\cellcolor[HTML]{FBE2D5}No}       & \multicolumn{1}{c|}{\cellcolor[HTML]{FBE2D5}No}                                                                 & \multicolumn{1}{c|}{\cellcolor[HTML]{FBE2D5}No}                                                                 & \multicolumn{1}{c|}{\cellcolor[HTML]{B5E6A2}Yes}                                                                     & \cellcolor[HTML]{B5E6A2}Yes                                                                     \\ \hline
\multicolumn{1}{|c|}{}                                      & \multicolumn{1}{c|}{}                                  & Hidden Layer Neurons             & \multicolumn{6}{c|}{\cellcolor[HTML]{EFEFEF}\# of features}                                                                                                                                                                                                                                                                                                                                                                                                                                                                                                                \\ \cline{3-9} 
\multicolumn{1}{|c|}{}                                      & \multicolumn{1}{c|}{}                                  & Output Layer Neurons             & \multicolumn{1}{c|}{\cellcolor[HTML]{EFEFEF}1}        & \multicolumn{1}{c|}{\cellcolor[HTML]{EFEFEF}1}        & \multicolumn{1}{c|}{\cellcolor[HTML]{EFEFEF}2}                                                                  & \multicolumn{1}{c|}{\cellcolor[HTML]{EFEFEF}2}                                                                  & \multicolumn{1}{c|}{\cellcolor[HTML]{EFEFEF}\# of classes}                                                           & \cellcolor[HTML]{EFEFEF}\# of classes                                                           \\ \cline{3-9} 
\multicolumn{1}{|c|}{}                                      & \multicolumn{1}{c|}{}                                  & Output Layer Activation          & \multicolumn{1}{c|}{\cellcolor[HTML]{E8E8E8}'linear'} & \multicolumn{1}{c|}{\cellcolor[HTML]{E8E8E8}'linear'} & \multicolumn{1}{c|}{\cellcolor[HTML]{E8E8E8}'sigmoid'}                                                          & \multicolumn{1}{c|}{\cellcolor[HTML]{E8E8E8}'sigmoid'}                                                          & \multicolumn{1}{c|}{\cellcolor[HTML]{E8E8E8}'softmax'}                                                               & \cellcolor[HTML]{E8E8E8}'softmax'                                                               \\ \cline{3-9} 
\multicolumn{1}{|c|}{}                                      & \multicolumn{1}{c|}{}                                  & Loss Function                    & \multicolumn{1}{c|}{\cellcolor[HTML]{E8E8E8}'mse'}    & \multicolumn{1}{c|}{\cellcolor[HTML]{E8E8E8}'mse'}    & \multicolumn{1}{c|}{\cellcolor[HTML]{E8E8E8}\begin{tabular}[c]{@{}c@{}}'binary\_\\  crossentropy'\end{tabular}} & \multicolumn{1}{c|}{\cellcolor[HTML]{E8E8E8}\begin{tabular}[c]{@{}c@{}}'binary\_\\  crossentropy'\end{tabular}} & \multicolumn{1}{c|}{\cellcolor[HTML]{E8E8E8}\begin{tabular}[c]{@{}c@{}}'categorical\_\\  crossentropy'\end{tabular}} & \cellcolor[HTML]{E8E8E8}\begin{tabular}[c]{@{}c@{}}'categorical\_\\  crossentropy'\end{tabular} \\ \cline{3-9} 
\multicolumn{1}{|c|}{}                                      & \multicolumn{1}{c|}{\multirow{-5}{*}{Hyperparameters}} & Metrics                          & \multicolumn{1}{c|}{\cellcolor[HTML]{E8E8E8}'mae'}    & \multicolumn{1}{c|}{\cellcolor[HTML]{E8E8E8}'mae'}    & \multicolumn{1}{c|}{\cellcolor[HTML]{E8E8E8}'accuracy'}                                                         & \multicolumn{1}{c|}{\cellcolor[HTML]{E8E8E8}'accuracy'}                                                         & \multicolumn{1}{c|}{\cellcolor[HTML]{E8E8E8}'accuracy'}                                                              & \cellcolor[HTML]{E8E8E8}'accuracy'                                                              \\ \cline{2-9} 
\multicolumn{1}{|c|}{}                                      & \multicolumn{1}{c|}{}                                  & FCNN Model                       & \multicolumn{1}{c|}{\cellcolor[HTML]{C0E4F5}\cmark}    & \multicolumn{1}{c|}{}                                 & \multicolumn{1}{c|}{\cellcolor[HTML]{C0E4F5}\cmark}                                                              & \multicolumn{1}{c|}{}                                                                                           & \multicolumn{1}{c|}{\cellcolor[HTML]{C0E4F5}\cmark}                                                                   &                                                                                                 \\ \cline{3-9} 
\multicolumn{1}{|c|}{\multirow{-7}{*}{\multirow{-2}{*}{\rotatebox{90}{\textbf{Actions}}}}}    & \multicolumn{1}{c|}{\multirow{-2}{*}{Mock Model}}      & CNN Model                        & \multicolumn{1}{c|}{}                                 & \multicolumn{1}{c|}{\cellcolor[HTML]{C0E4F5}\cmark}    & \multicolumn{1}{c|}{}                                                                                           & \multicolumn{1}{c|}{\cellcolor[HTML]{C0E4F5}\cmark}                                                              & \multicolumn{1}{c|}{}                                                                                                & \cellcolor[HTML]{C0E4F5}\cmark                                                                   \\ \hline

\end{tabular}

\end{adjustbox}
}
\label{tab:decisiontablemodel}
\vspace{-0.15cm}
\end{table}

\subsection{Mock Object Creation and Verification}
To ensure the correctness of each module, it is essential to verify 
that each module exhibits the correct functionality depending on the task.
Our insight is that to evaluate the expected behavior of each module, useful mocks can be constructed that approximate the behavior of the original data or model using the information exposed in interfaces.
In the mock implementation, the primary goal is to achieve simplicity instead of aiming for completeness~\cite{mackinnon2000endotesting}.
To that end, we propose a systematic approach for creating mocks for the data preparation and model design stages, detailing how these mocks are utilized for verification.
Below we discuss 
the process for each stage in detail.

\subsubsection{Data Preparation Stage}
The data preparation module intends to produce good-quality data and involves various tasks such as 
data cleaning, handling missing values, \etc. 
Typically, the features are selected and refined based on the model's performance during evaluation, establishing a feedback loop from the model evaluation stage to the data preparation stage~\cite{biswas22art}.
Our insight is that by creating a mock model that approximates the behavior of the original model, data quality can be assessed and enhanced through early-stage evaluation.
While the mock model doesn't have to preserve every semantic detail, 
it is essential to generate a model with the appropriate hyperparameters that yield the correct output tailored to a task.
The mock model can be automatically generated using the interface of the model design stage
as illustrated in \figref{fig:MockCreation}(a). 
Below, we discuss the process of mock model creation.

\paragraph{Process for Mock Model Creation} Generating a mock model involves several key steps to ensure the correctness and reliability of the generated model.

\textbf{Adaptive Mock Model Generation:} DL model has a lot of hyperparameters, which are provided at the time of model design by the developer.
The choice of the hyperparameters depends on several factors, such as the type of task and complexity of the dataset~\cite{yu2020hyper,LeCun98LeNet5}.
Any incorrect hyperparameter can be misleading, giving rise to bugs due to inaccuracies in the model.
Therefore, for automatic mock model generation,
it is necessary to construct adaptive mocks that change based on the task at hand and adapt to different testing conditions without manual intervention.
This adaptability ensures that the mock model aligns with testing requirements. Furthermore, as this paper introduces a design paradigm that supports independent development, developers can utilize automatically generated mock models for testing the data without delving into the intricacies of designing them.

Our approach for the automatic mock model generation is described in Decision Table~\ref{tab:decisiontablemodel}.
For initializing the mock model's hyperparameters, we reviewed the AI literature~\cite{Krizhevsky12AlexNet,lecun2002efficient,bengio2012practical} and Keras documentation~\cite{Kerasloss}. 
We utilized the hyperparameters suggested by the literature for a given task.
For example, for the DL program  in~\figref{Motivation}, the conditions as outlined in Decision Table~\ref{tab:decisiontablemodel} are problem type - `regression', model type - `FCNN', and classes - 1 (set to 1 for regression). The corresponding actions generate a mock model with hidden layer neurons equal to the `\# of features', an output layer with 1 neuron and a `linear' activation function, and a compilation layer with the `mse' loss function and `mae' as the metrics.

\textbf{Complexity of Mock Model:}
DL models are usually complex with several parameters and their complexity increases with the type of task at hand.
Determining the complexity of the model requires careful consideration during mock model generation.
Since the mock model aims to verify the quality of the data, creating a complex mock model for the unit testing could lead to excessive resource usage without serving the primary goal of unit testing.

To optimize the model's complexity for unit testing while managing resource consumption, we propose creating a mock model 
consisting of only three layers: the input layer, one hidden layer, and the output layer.
The rationale for opting for the simplest network is influenced by the 
principle of ``Start Simple'', as recommended in the machine learning literature~\cite{andrewng, occamrazor}. 
If a simple network struggles to learn from the training data, it suggests that the training data requires further refinement~\cite{yu2020hyper}.
Next, we explain how this mock model is utilized for the verification of data.

\paragraph{Verification of Data using Mock Model}
After generating a mock model that simulates the behavior of the original DNN model, verification is performed by inputting the preprocessed data into the mock model, rather than using the original model. Certain data preparation issues, like missing data, can be detected through data property assertions. In contrast, issues like mislabeled data, outliers, or improper feature selection require in-depth analysis. 
These issues often manifest as subtle errors that impact model performance and require a thorough examination of data-model interactions for effective identification. Therefore, we propose an integrated approach that combines data property assertions with analysis of the mock model's behavior on preprocessed data. The data property assertions are used to identify fundamental issues (1-6 in the data preparation stage shown in \figref{Overview}).  
Moreover, more complex issues 
labeled as ‘Model not Learning’ in \figref{Overview} are identified by observing the mock model’s behavior. Symptoms such as high loss, frequent misclassifications, or consistently low confidence on specific samples point to these issues, highlighting areas that require further investigation. This approach allows for the verification and refinement of data before it is used to train the original model.

\subsubsection{Model Design Stage}
The goal of the model design stage is to generate a model with suitable hyperparameters and correct API usage,
appropriate for the given task. This facilitates the model in learning features from the training data. 
Since DNNs are data-driven, training data produced by the data preparation stage is typically used to evaluate the model's performance and tune its hyperparameters. 
Although the mock data might not help detect all the training time issues,
our insight is that leveraging mock data enables the early detection of numerous bugs, including tensor shape mismatches, inappropriate hyperparameter selection, \etc. It allows for improving the model's quality before assessing its performance on original training data.
\figref{fig:MockCreation}(b) illustrates the mock data creation process using the interface of the data preparation stage.
While the mock data doesn’t have to preserve every semantic detail, it is crucial to ensure that it does not contain missing values or outliers.
Below, we discuss the process of mock data generation.

\paragraph{Process for Mock Data Creation}
Generating mock data requires careful adherence to key steps to ensure the correctness of the resulting data.

\textbf{Preprocessed Mock Data Generation:}
During the data preparation stage, several preprocessing steps such as data cleaning, outlier removal, class balancing, \etc., are performed to ensure the quality of the data.
Therefore, while creating mock data, it's crucial to include similar preprocessing steps to avoid errors from inaccuracies in the mock data.

To ensure that the mock data accurately reflects the real-world scenarios, we utilized the \textit{make\_classification()} and \textit{make\_regression()} functions provided by \textit{scikit-learn}~\cite{scikit} for synthetic data generation. 
The advantage of utilizing these functions is that these functions provide a level of control over the characteristics of the generated dataset.
For example, we can generate mock data that is normally distributed, without outliers, is labeled and classes are balanced. 
These functions are also customizable, allowing users to specify the number of samples, features, \etc., offering flexibility for specific testing scenarios.
Although the data generated by these functions is normally distributed, however, it is not scaled or normalized.
Scaling or normalization is a common preprocessing step that aids in faster convergence of DL models during training~\cite{LeCun12}.
In our approach to mock data generation, we scaled the data generated 
by \textit{make\_classification()} and \textit{make\_regression()} functions.
The scaled mock data is then utilized to verify the behavior of the model. 
The rationale for validating the model's behavior with mock data is that if the model struggles to learn from the mock data,
the model's hyperparameters can be refined before proceeding to train it with the original dataset, which is usually more complex than the mock data.

\textbf{Quantification of Mock Data:}
\label{quantificationdata}
Determining the amount of data necessary for training a DL model is often a subject of debate.
It is typically gauged by several factors, such as the complexity of the dataset and the model's performance during evaluation.
In this paper, mock data is employed for unit testing to ensure the precise functioning of the mathematical functions within each layer of the designed model. It also verifies the transformation of input data into meaningful representations tailored to the specific task for which the model is designed.
Therefore, determining the amount of data required for the intended purpose (\ie, unit testing) poses a challenge.
A small number of samples can lead to misleading results, while a large volume of samples can be resource-intensive.

To address this challenge, we consulted established machine learning literature~\cite{ruleof10,lakshmanan2020machine} to estimate the approximate dataset size. We performed a sensitivity analysis of the data set sizes suggested in the literature~\cite{ruleof10,lakshmanan2020machine}.
We varied the dataset size in increments $\pm5\%$, $\pm10\%$ and $\pm20\%$ and observed the impact on the model's performance. 
Based on this analysis, the samples generated by {\em KUnit} are as follows:
for the regression task, the number of samples generated is 10 times the number of features and for the classification task, 100 samples are generated for each class.
In our experiments, we found that these samples were adequate for identifying various types of issues in the designed models.
We now discuss how the mock data is utilized for the verification of the model.

\paragraph{Verification of Model using Mock Data}
After generating mock data that mimics key characteristics of the original dataset, verification is done by feeding the mock data into the designed model. This ensures the model's correctness without depending on the original data. To facilitate this, we propose an integrated approach that combines the model property assertions with an analysis of the model's behavior on mock data. For verifying the model’s 
structure,
assertions are defined using the data properties defined in the interface of the data preparation stage and conditions obtained from the literature
(Section~\ref{bugconditions}), which helps to identify issues 
A-E illustrated in \figref{Overview} for the model design stage. Next, the model's response to the mock data is analyzed to ensure it appropriately handles normalized/scaled data. This analysis facilitates the early detection and resolution of potential issues F-G, shown in \figref{Overview}.
By detecting these issues early, the approach allows verification and refinement of the 
model structure before using the original data for training.

\section{Evaluation}\label{sec:EVALUATION}

\subsection{Experimental Setup}
In this section, we discuss the process for collecting assertions for issue identification, datasets used for the empirical evaluation and user study, task description, and details of participants involved in the user study.

\subsubsection{Procedure for Collecting Assertions for Issue Identification}\label{bugconditions}

In this section, we detail our process for identifying the types of bugs supported by {\em KUnit} and explain how the corresponding assertions are developed to detect them. To identify these bugs, we conducted a thorough literature review. 
Islam \etal~\cite{islam19} investigated the type of DL bugs and categorized them into different categories, with data and model bugs being two key categories. Humbatova \etal~\cite{humbatova20taxonomy} refined the investigation and further divided data bugs into two subcategories:
training data quality and preprocessing of training data. And, the model bugs were categorized into subcategories such as wrong input, wrong tensor shape, \etc. These classifications provided a structured foundation for understanding common pitfalls in DL workflow. While some bugs reported in these studies require comprehensive end-to-end analysis, others can be effectively detected through targeted testing of specific components, such as data and model. For instance, focused component-level testing can detect crash bugs caused by wrong preprocessing and silent bugs resulting from incorrect activation functions. In contrast, issues like overfitting and underfitting depend on evaluating the model's performance on the original dataset. Thus, we focus on bugs that can be detected at the component level while excluding those that require end-to-end analysis. Similar to the procedure described in \cite{BraiekDeepChecker,ghanbari2023deepmufl}, we filtered out these bugs from the empirical studies~\cite{islam19,humbatova20taxonomy}  and obtained 7 data-related issues (1-7 in Fig. 2) and 8 model-design-related issues (A-H in Fig. 2) currently supported by {\em KUnit}. 
We then adopted an approach similar to that of TheDeepChecker~\cite{BraiekDeepChecker} and reviewed existing works on fault localization and repair
techniques for DL programs~\cite{wardat21DeepLocalize, wardat22DeepDiagnosis, cao2022deepfd, Zhang21Autotrainer, nikanjam2021neuralint, manke2024leveraging, wardat2023localizing, islam20repairing}, contracts for DL programs~\cite{ahmed23dlcontract, Khairunnesa2023}, and the Keras official documentation~\cite{Keras,kerasexamples}. This review enhanced our understanding of the root cause of the bugs and how these issues manifest in DL workflow, allowing us to establish the conditions necessary to identify and address them. These conditions are implemented as assertions in {\em KUnit}'s test methods to identify the bugs and repair strategies are utilized to provide actionable fixes in {\em KUnit}.

\subsubsection{Implementation}
We implemented {\em KUnit} in Python on top of \textit{Keras} 2.3.0 and \textit{TensorFlow} 2.1.0. 
The conditions obtained in Section~\ref{bugconditions} are implemented as test cases using Python's built-in unit testing framework, \texttt{unittest}.

\subsubsection{Empirical Evaluation}
To evaluate our approach, we collected DL programs developed using Keras. 
We examined the recently published DL fault localization benchmarks~\cite{wardat21DeepLocalize,cao2022deepfd,ghanbari2023deepmufl}.
Currently, {\em KUnit}, supports two types of DL architectures, FCNNs and CNNs designed for regression and classification problems for structural data. 
We used these criteria for filtering the programs from these benchmarks.
We identified an overlap among the programs in these benchmarks and filtered out duplicates to retain only unique programs. As a result, we obtained 50 programs, 42 from \textit{Stack Overflow} and 8 from \textit{GitHub}.
We considered the 50 programs in our benchmark as ``unseen''
because we have not seen these buggy and correct programs during the process of acquiring 
conditions described in Section~\ref{bugconditions}.

\subsubsection{User Study}

We also performed a user study to evaluate our approach.
We followed
the methodology of Biswas \etal~\cite{biswas22art} to collect real-world datasets and tasks from Kaggle competitions~\cite{kaggle} and collected 5 real-world datasets that require preprocessing to meet data quality requirements. 
The details of these datasets are shown in Table~\ref{tab:userstudydatasets}. 
We selected 5 sequential DL models from Kaggle competitions~\cite{kaggle} and provided them as a reference model to the participants. 
The details of these models are shown in Table~\ref{tab:userstudymodels}.

\begin{table}[]
\caption{Datasets used for user study.}
 \centering
\scalebox{0.65}{
\begin{tabular}{ccccccc}
\hline
\rowcolor[HTML]{C0C0C0}
\multicolumn{2}{|c|}{\textbf{Datasets}}                                                                                            & \multicolumn{1}{c|}{\textbf{\begin{tabular}[c]{@{}c@{}}Portfolio\\  Data = NU \\  Labels = NU\\  Regression\end{tabular}}} & \multicolumn{1}{c|}{\textbf{\begin{tabular}[c]{@{}c@{}}Grain\\  Data = NU\\  Labels = CA\\  Multiclass\\  Classification\end{tabular}}} & \multicolumn{1}{c|}{\textbf{\begin{tabular}[c]{@{}c@{}}Truck\\  Data = MI\\  Labels = NU\\  Regression\end{tabular}}} & \multicolumn{1}{c|}{\textbf{\begin{tabular}[c]{@{}c@{}}Loan\\  Data = MI \\  Labels = NU\\  Binary\\  Classification\end{tabular}}} & \multicolumn{1}{c|}{\textbf{\begin{tabular}[c]{@{}c@{}}Train\\  Data = MI\\  Labels = NU\\  Binary\\  Classification\end{tabular}}} \\ \hline
\multicolumn{2}{|c|}{\textbf{Data Checks}}                                                                                         & \multicolumn{5}{c|}{\textbf{Issues}}                                                                                                                                                                                                                                                                                                                                                                                                                                                                                                                                                                                                                                       \\ \hline
\multicolumn{1}{|c|}{\multirow{3}{*}{\textbf{Data Quality}}}                                                      & \multicolumn{1}{c|}{MV} & \multicolumn{1}{c|}{N}                                                                                                     & \multicolumn{1}{c|}{N}                                                                                                                  & \multicolumn{1}{c|}{Y}                                                                                                & \multicolumn{1}{c|}{Y}                                                                                                              & \multicolumn{1}{c|}{Y}                                                                                                              \\ \cline{2-7} 
\multicolumn{1}{|c|}{}                                                                                   & \multicolumn{1}{c|}{ML} & \multicolumn{1}{c|}{N}                                                                                                     & \multicolumn{1}{c|}{N}                                                                                                                  & \multicolumn{1}{c|}{N}                                                                                                & \multicolumn{1}{c|}{N}                                                                                                              & \multicolumn{1}{c|}{Y}                                                                                                              \\ \cline{2-7} 
\multicolumn{1}{|c|}{}                                                                                   & \multicolumn{1}{c|}{CI} & \multicolumn{1}{c|}{N}                                                                                                     & \multicolumn{1}{c|}{Y}                                                                                                                  & \multicolumn{1}{c|}{N}                                                                                                & \multicolumn{1}{c|}{Y}                                                                                                              & \multicolumn{1}{c|}{N}                                                                                                              \\ \hline
\multicolumn{1}{|c|}{\multirow{2}{*}{\begin{tabular}[c]{@{}c@{}}\textbf{Preprocessing} \\  \textbf{of Data}\end{tabular}}} & \multicolumn{1}{c|}{ME} & \multicolumn{1}{c|}{N}                                                                                                     & \multicolumn{1}{c|}{Y}                                                                                                                  & \multicolumn{1}{c|}{Y}                                                                                                & \multicolumn{1}{c|}{Y}                                                                                                              & \multicolumn{1}{c|}{Y}                                                                                                              \\ \cline{2-7} 
\multicolumn{1}{|c|}{}                                                                                   & \multicolumn{1}{c|}{MS} & \multicolumn{1}{c|}{Y}                                                                                                     & \multicolumn{1}{c|}{Y}                                                                                                                  & \multicolumn{1}{c|}{Y}                                                                                                & \multicolumn{1}{c|}{Y}                                                                                                              & \multicolumn{1}{c|}{Y}                                                                                                              \\ \hline
\multicolumn{7}{c}{\begin{tabular}[c]{@{}c@{}}MV = Missing/Infinite Value, ML = Missing Label, CI = Class Imbalance, \\  ME = Missing encoding of categorical data, MS = Missing Scaling/Normalization\end{tabular}}                                                                                                                                                                                                                                                                                                                                                                                                                                                                                                                                                                                          \\
\multicolumn{7}{c}{NU = Numeric, CA = Categorical, MI = Mixed}                                                                                                                                                                                                                                                                                                                                                                                                                                                                                     
\end{tabular}
\label{tab:userstudydatasets}
}
\vspace{-0.3cm}
\end{table}

\begin{table}[]
\caption{Models used for user study.}
 \centering
\scalebox{0.85}{
\begin{tabular}{|c|c|c|c|c|}
\hline
\rowcolor[HTML]{C0C0C0} 
\textbf{Model \#} & \textbf{\begin{tabular}[c]{@{}c@{}}Architecture \\  Type\end{tabular}} & \textbf{\# of layers} & \textbf{\# of neurons} & \textbf{\# of parameters} \\ \hline
M1                & FCNN                                                                   & 5                     & 21                     & 120                       \\ \hline
M2                & FCNN                                                                   & 6                     & 131                    & 5703                      \\ \hline
M3                & FCNN                                                                   & 3                     & 41                     & 601                       \\ \hline
M4                & FCNN                                                                   & 6                     & 57                     & 1153                      \\ \hline
M5                & CNN                                                                    & 5                     & 71                     & 5201                      \\ \hline
\end{tabular}
\label{tab:userstudymodels}
}
\vspace{-0.2cm}
\end{table}
\subsubsection{Tasks}
To avoid versioning issues during experimental setup and save participants time, we hosted tasks on \textit{GitHub} Codespaces~\cite{githubcodespaces}; a web-based VS Code IDE that allows developers to edit, run, test, and debug code within a web browser.
We used Zoom to monitor each participant's screen and ensured they used the program as intended.
We shared with participants a document with instructions explaining the goal of the task and how to run it on \textit{GitHub} Codespaces IDE.
In our task design, each participant performed two tasks: one using the traditional approach and the other using the modular approach.
Since participants working on a task are likely to remember its details and the issues encountered, we adopted a between-subjects design~\cite{davis2023nanofuzz}  to mitigate the learning effect. Specifically, we assigned each participant two distinct problems.
For the data preparation task, 
participants were provided with datasets and instructed to explore the data and apply preprocessing steps based on their expertise.
In the traditional setting, they did not test the code, whereas in the modular setting, participants tested the added preprocessing steps in isolation using an automatically generated mock model. 
For the model designing task, participants were given a reference DL model and asked to modify its structure based on the task requirements and their experience. In the traditional setting, participants received preprocessed data from the data preparation task completed by another participant in the traditional setting and tested the designed model using the original data. Whereas, in a modular setting, participants tested the designed model in isolation using automatically generated mock data.
During the study, participants were allowed to access the internet to verify the syntax of different operations in Python and Keras. 
After completion of the task, we asked participants to complete a survey to share their experience of {\em KUnit} on a 5-point Likert scale and provide open-ended feedback.
We conducted a pilot study with 7 participants, which allowed us to refine the tasks and instructions. This study was reviewed by our Institutional Review Board.

\begin{table}[]
\centering
\caption{Summary of issues detected by {\em KUnit}.}

\scalebox{0.8}{

\begin{tabular}{|cl|c|c|}
\hline
\rowcolor[HTML]{C0C0C0} 
\multicolumn{1}{|c|}{\cellcolor[HTML]{C0C0C0}\textbf{Stage}}                                         & \multicolumn{1}{c|}{\cellcolor[HTML]{C0C0C0}\textbf{Categories}} & \textbf{\begin{tabular}[c]{@{}c@{}}\# of issues \\ in different\\  categories\end{tabular}} & \textbf{\begin{tabular}[c]{@{}c@{}}\# of issues \\  detected by \\ KUnit\end{tabular}} \\ \hline
\multicolumn{1}{|c|}{}                                                                               & Missing Scaling/Normalization                                    & 11                                                                                          & 9                                                                                      \\ \cline{2-4} 
\multicolumn{1}{|c|}{\multirow{-2}{*}{\begin{tabular}[c]{@{}c@{}}\textbf{Data} \\  \textbf{Preparation}\end{tabular}}} & Labels not matching problem definition                           & 1                                                                                           & 1                                                                                      \\ \hline
\multicolumn{1}{|c|}{}                                                                               & Incorrect Input shape                                            & 1                                                                                           & 1                                                                                      \\ \cline{2-4} 
\multicolumn{1}{|c|}{}                                                                               & Incorrect Output shape                                           & 2                                                                                           & 2                                                                                      \\ \cline{2-4} 
\multicolumn{1}{|c|}{}                                                                               & Missing Activations                                              & 2                                                                                           & 2                                                                                      \\ \cline{2-4} 
\multicolumn{1}{|c|}{}                                                                               & Wrong Output Layer Activation                                    & 30                                                                                          & 27                                                                                     \\ \cline{2-4} 
\multicolumn{1}{|c|}{}                                                                               & Learning Rate out of Common Range                                & 3                                                                                           & 3                                                                                      \\ \cline{2-4} 
\multicolumn{1}{|c|}{}                                                                               & Wrong Loss Function                                              & 8                                                                                           & 8                                                                                      \\ \cline{2-4} 
\multicolumn{1}{|c|}{}                                                                               & Incorrect Evaluation Metrics                                     & 7                                                                                           & 7                                                                                      \\ \cline{2-4} 
\multicolumn{1}{|c|}{\multirow{-8}{*}{\begin{tabular}[c]{@{}c@{}}\textbf{Model}\\   \textbf{Design}\end{tabular}}}  & Oscillating Loss/Slow Convergence                                & 9                                                                                           & 3                                                                                      \\ \hline
\multicolumn{2}{|c|}{\textbf{Total}}                                                                                                                                    & \textbf{74}                                                                                 & \textbf{63}                                                                            \\ \hline

\end{tabular}
\label{SOFsummary}
}
\vspace{-0.15cm}
\end{table}

\begin{table*}[]
\huge
\caption{Summary of issues detected by {\em KUnit} using mock model in data preparation stage.}
\renewcommand{\arraystretch}{1}
\begin{adjustbox}{width=\textwidth,center}

\scalebox{0.7}{
\begin{tabular}{|cll|ccc|ccc|ccc|ccc|ccc|}
\hline
\rowcolor[HTML]{BFBFBF} 
\multicolumn{1}{|c|}{\cellcolor[HTML]{BFBFBF}}                                                                                    & \multicolumn{2}{c|}{\cellcolor[HTML]{BFBFBF}}                                                                                                                                                                   & \multicolumn{3}{c|}{\cellcolor[HTML]{BFBFBF}\textbf{Task 1 (Portfolio)}}                                                                                                                                                                                                               & \multicolumn{3}{c|}{\cellcolor[HTML]{BFBFBF}\textbf{Task 2 (Grain)}}                                                                                                                                                                                                                   & \multicolumn{3}{c|}{\cellcolor[HTML]{BFBFBF}\textbf{Task 3 (Truck)}}                                                                                                                                                                                                                  & \multicolumn{3}{c|}{\cellcolor[HTML]{BFBFBF}\textbf{Task 4 (Loan)}}                                                                                                                                                                                                                   & \multicolumn{3}{c|}{\cellcolor[HTML]{BFBFBF}\textbf{Task 5 (Train)}}                                                                                                                                                                                                                  \\ \cline{4-18} 
\rowcolor[HTML]{BFBFBF} 
\multicolumn{1}{|c|}{\multirow{-2}{*}{\cellcolor[HTML]{BFBFBF}\textbf{Stage}}}                                                    & \multicolumn{2}{c|}{\multirow{-2}{*}{\cellcolor[HTML]{BFBFBF}\textbf{Tests}}}                                                                                                                                   & \multicolumn{1}{c|}{\cellcolor[HTML]{BFBFBF}\textbf{\begin{tabular}[c]{@{}c@{}}S1 \\  (Com)\end{tabular}}} & \multicolumn{1}{c|}{\cellcolor[HTML]{BFBFBF}\textbf{\begin{tabular}[c]{@{}c@{}}S2\\  (Com)\end{tabular}}} & \textbf{\begin{tabular}[c]{@{}c@{}}S3\\  (Prof)\end{tabular}} & \multicolumn{1}{c|}{\cellcolor[HTML]{BFBFBF}\textbf{\begin{tabular}[c]{@{}c@{}}S1\\  (Prof)\end{tabular}}} & \multicolumn{1}{c|}{\cellcolor[HTML]{BFBFBF}\textbf{\begin{tabular}[c]{@{}c@{}}S2\\  (Prof)\end{tabular}}} & \textbf{\begin{tabular}[c]{@{}c@{}}S3\\  (Exp)\end{tabular}} & \multicolumn{1}{c|}{\cellcolor[HTML]{BFBFBF}\textbf{\begin{tabular}[c]{@{}c@{}}S1\\  (Com)\end{tabular}}} & \multicolumn{1}{c|}{\cellcolor[HTML]{BFBFBF}\textbf{\begin{tabular}[c]{@{}c@{}}S2\\  (Prof)\end{tabular}}} & \textbf{\begin{tabular}[c]{@{}c@{}}S3\\  (Exp)\end{tabular}} & \multicolumn{1}{c|}{\cellcolor[HTML]{BFBFBF}\textbf{\begin{tabular}[c]{@{}c@{}}S1\\  (Com)\end{tabular}}} & \multicolumn{1}{c|}{\cellcolor[HTML]{BFBFBF}\textbf{\begin{tabular}[c]{@{}c@{}}S2\\  (Prof)\end{tabular}}} & \textbf{\begin{tabular}[c]{@{}c@{}}S3\\  (Exp)\end{tabular}} & \multicolumn{1}{c|}{\cellcolor[HTML]{BFBFBF}\textbf{\begin{tabular}[c]{@{}c@{}}S1\\  (Com)\end{tabular}}} & \multicolumn{1}{c|}{\cellcolor[HTML]{BFBFBF}\textbf{\begin{tabular}[c]{@{}c@{}}S2\\  (Prof)\end{tabular}}} & \textbf{\begin{tabular}[c]{@{}c@{}}S3\\  (Exp)\end{tabular}} \\ \hline
\multicolumn{1}{|c|}{}                                                                                                            & \multicolumn{1}{l|}{\begin{tabular}[c]{@{}l@{}}Preprocessing steps are \\  applied correctly\end{tabular}} & \begin{tabular}[c]{@{}l@{}}Check scaling/normalization \\  is done correctly\end{tabular}          & \multicolumn{1}{c|}{\cellcolor[HTML]{DAF2D0}\xmark}                                                             & \multicolumn{1}{c|}{\cellcolor[HTML]{DAF2D0}\xmark}                                                            & \cellcolor[HTML]{DAF2D0}\xmark                                     & \multicolumn{1}{c|}{\cmark}                                                                                     & \multicolumn{1}{c|}{\cmark}                                                                                     & \cmark                                                            & \multicolumn{1}{c|}{\cmark}                                                                                    & \multicolumn{1}{c|}{\cmark}                                                                                     & \cellcolor[HTML]{DAF2D0}\xmark                                    & \multicolumn{1}{c|}{\cellcolor[HTML]{DAF2D0}\xmark}                                                            & \multicolumn{1}{c|}{\cmark}                                                                                     & \cmark                                                            & \multicolumn{1}{c|}{\cmark}                                                                                    & \multicolumn{1}{c|}{\cmark}                                                                                     & \cellcolor[HTML]{DAF2D0}\xmark                                    \\ \cline{2-18} 
\multicolumn{1}{|c|}{}                                                                                                            & \multicolumn{1}{l|}{}                                                                                      & Missing values are removed/replaced                                                                & \multicolumn{1}{c|}{--}                                                                                    & \multicolumn{1}{c|}{--}                                                                                   & --                                                            & \multicolumn{1}{c|}{--}                                                                                    & \multicolumn{1}{c|}{--}                                                                                    & --                                                           & \multicolumn{1}{c|}{\cellcolor[HTML]{DAF2D0}\xmark}                                                            & \multicolumn{1}{c|}{\cellcolor[HTML]{DAF2D0}\xmark}                                                             & \cellcolor[HTML]{DAF2D0}\xmark                                    & \multicolumn{1}{c|}{\cmark}                                                                                    & \multicolumn{1}{c|}{\cmark}                                                                                     & \cmark                                                            & \multicolumn{1}{c|}{\cmark}                                                                                    & \multicolumn{1}{c|}{\cmark}                                                                                     & \cmark                                                            \\ \cline{3-18} 
\multicolumn{1}{|c|}{}                                                                                                            & \multicolumn{1}{l|}{}                                                                                      & Missing label are removed/replaced                                                                 & \multicolumn{1}{c|}{--}                                                                                    & \multicolumn{1}{c|}{--}                                                                                   & --                                                            & \multicolumn{1}{c|}{--}                                                                                    & \multicolumn{1}{c|}{--}                                                                                    & --                                                           & \multicolumn{1}{c|}{--}                                                                                   & \multicolumn{1}{c|}{--}                                                                                    & --                                                           & \multicolumn{1}{c|}{--}                                                                                   & \multicolumn{1}{c|}{--}                                                                                    & --                                                           & \multicolumn{1}{c|}{\cellcolor[HTML]{DAF2D0}\xmark}                                                            & \multicolumn{1}{c|}{\cellcolor[HTML]{DAF2D0}\xmark}                                                             & \cmark                                                            \\ \cline{3-18} 
\multicolumn{1}{|c|}{}                                                                                                            & \multicolumn{1}{l|}{}                                                                                      & Classes are balanced                                                                               & \multicolumn{1}{c|}{--}                                                                                    & \multicolumn{1}{c|}{--}                                                                                   & --                                                            & \multicolumn{1}{c|}{\cellcolor[HTML]{DAF2D0}\xmark}                                                             & \multicolumn{1}{c|}{\cellcolor[HTML]{DAF2D0}\xmark}                                                             & \cellcolor[HTML]{DAF2D0}\xmark                                    & \multicolumn{1}{c|}{--}                                                                                   & \multicolumn{1}{c|}{--}                                                                                    & --                                                           & \multicolumn{1}{c|}{\cellcolor[HTML]{DAF2D0}\xmark}                                                            & \multicolumn{1}{c|}{\cellcolor[HTML]{DAF2D0}\xmark}                                                             & \cmark                                                            & \multicolumn{1}{c|}{\cellcolor[HTML]{DAF2D0}\xmark}                                                            & \multicolumn{1}{c|}{\cellcolor[HTML]{DAF2D0}\xmark}                                                             & \cmark                                                            \\ \cline{3-18} 
\multicolumn{1}{|c|}{}                                                                                                            & \multicolumn{1}{l|}{\multirow{-4}{*}{Quality assurance}}                                                   & Labels are matching problem definition                                                             & \multicolumn{1}{c|}{--}                                                                                    & \multicolumn{1}{c|}{--}                                                                                   & --                                                            & \multicolumn{1}{c|}{--}                                                                                    & \multicolumn{1}{c|}{--}                                                                                    & --                                                           & \multicolumn{1}{c|}{--}                                                                                   & \multicolumn{1}{c|}{--}                                                                                    & --                                                           & \multicolumn{1}{c|}{--}                                                                                   & \multicolumn{1}{c|}{--}                                                                                    & --                                                           & \multicolumn{1}{c|}{\cellcolor[HTML]{DAF2D0}\xmark}                                                            & \multicolumn{1}{c|}{\cellcolor[HTML]{DAF2D0}\xmark}                                                             & \cmark                                                            \\ \cline{2-18} 
\multicolumn{1}{|c|}{}                                                                                                            & \multicolumn{1}{l|}{Correct format of data}                                                                & \begin{tabular}[c]{@{}l@{}}Categorical data is converted to \\  numeric data\end{tabular}          & \multicolumn{1}{c|}{--}                                                                                    & \multicolumn{1}{c|}{--}                                                                                   & --                                                            & \multicolumn{1}{c|}{\cmark}                                                                                     & \multicolumn{1}{c|}{\cmark}                                                                                     & \cmark                                                            & \multicolumn{1}{c|}{\cmark}                                                                                    & \multicolumn{1}{c|}{\cmark}                                                                                     & \cmark                                                            & \multicolumn{1}{c|}{\cmark}                                                                                    & \multicolumn{1}{c|}{\cmark}                                                                                     & \cmark                                                            & \multicolumn{1}{c|}{\cmark}                                                                                    & \multicolumn{1}{c|}{\cmark}                                                                                     & \cmark                                                            \\ \cline{2-18} 
\multicolumn{1}{|c|}{\multirow{-7}{*}{\begin{tabular}[c]{@{}c@{}}Data Preparation\\  (Testing using \\ \textbf{Mock Model})\end{tabular}}} & \multicolumn{1}{l|}{\begin{tabular}[c]{@{}l@{}}Performance of the \\ mock model\end{tabular}}              & \begin{tabular}[c]{@{}l@{}}Check mock model is able to learn\\ from selected features\end{tabular} & \multicolumn{1}{c|}{\cmark}                                                                                     & \multicolumn{1}{c|}{\cmark}                                                                                    & \cmark                                                             & \multicolumn{1}{c|}{\cmark}                                                                                     & \multicolumn{1}{c|}{\cmark}                                                                                     & \cmark                                                            & \multicolumn{1}{c|}{\cellcolor[HTML]{DAF2D0}\xmark}                                                            & \multicolumn{1}{c|}{\cellcolor[HTML]{DAF2D0}\xmark}                                                             & \cellcolor[HTML]{DAF2D0}\xmark                                    & \multicolumn{1}{c|}{\cmark}                                                                                    & \multicolumn{1}{c|}{\cmark}                                                                                     & \cmark                                                            & \multicolumn{1}{c|}{\cellcolor[HTML]{DAF2D0}\xmark}                                                            & \multicolumn{1}{c|}{\cellcolor[HTML]{DAF2D0}\xmark}                                                             & \cmark                                                            \\ \hline
\multicolumn{3}{|c|}{\textbf{Number of issues detected in data preprocessing}}                                                                                                                                                                                                                                                                      & \multicolumn{1}{c|}{\textbf{1}}                                                                            & \multicolumn{1}{c|}{\textbf{1}}                                                                           & \textbf{1}                                                    & \multicolumn{1}{c|}{\textbf{1}}                                                                            & \multicolumn{1}{c|}{\textbf{1}}                                                                            & \textbf{1}                                                   & \multicolumn{1}{c|}{\textbf{2}}                                                                           & \multicolumn{1}{c|}{\textbf{2}}                                                                            & \textbf{3}                                                   & \multicolumn{1}{c|}{\textbf{2}}                                                                           & \multicolumn{1}{c|}{\textbf{1}}                                                                            & \textbf{0}                                                   & \multicolumn{1}{c|}{\textbf{4}}                                                                           & \multicolumn{1}{c|}{\textbf{4}}                                                                            & \textbf{1}                                                   \\ \hline
\multicolumn{3}{|c|}{\textbf{Number of issues resolved by participants}}                                                                                                                                                                                                                                                                            & \multicolumn{1}{c|}{\textbf{1}}                                                                            & \multicolumn{1}{c|}{\textbf{1}}                                                                           & \textbf{1}                                                    & \multicolumn{1}{c|}{\textbf{1}}                                                                            & \multicolumn{1}{c|}{\textbf{1}}                                                                            & \textbf{1}                                                   & \multicolumn{1}{c|}{\textbf{2}}                                                                           & \multicolumn{1}{c|}{\textbf{2}}                                                                            & \textbf{3}                                                   & \multicolumn{1}{c|}{\textbf{2}}                                                                           & \multicolumn{1}{c|}{\textbf{1}}                                                                            & \textbf{0}                                                   & \multicolumn{1}{c|}{\textbf{4}}                                                                           & \multicolumn{1}{c|}{\textbf{4}}                                                                            & \textbf{1}                                                   \\ \hline
\multicolumn{3}{|c|}{\textbf{Number of issues dismissed as false alarms by participants}}                                                                                                                                                                                                                                                           & \multicolumn{1}{c|}{\textbf{0}}                                                                            & \multicolumn{1}{c|}{\textbf{0}}                                                                           & \textbf{0}                                                    & \multicolumn{1}{c|}{\textbf{0}}                                                                            & \multicolumn{1}{c|}{\textbf{0}}                                                                            & \textbf{0}                                                   & \multicolumn{1}{c|}{\textbf{0}}                                                                           & \multicolumn{1}{c|}{\textbf{0}}                                                                            & \textbf{0}                                                   & \multicolumn{1}{c|}{\textbf{0}}                                                                           & \multicolumn{1}{c|}{\textbf{0}}                                                                            & \textbf{0}                                                   & \multicolumn{1}{c|}{\textbf{0}}                                                                           & \multicolumn{1}{c|}{\textbf{0}}                                                                            & \textbf{0}                                                   \\ \hline
\end{tabular}
}
\label{rq1userstudydata}
\end{adjustbox}

\begin{tablenotes}
 \centering
        \fontsize{7.5pt}{7.5pt}\selectfont
        \item \textbf{Expertise level of participants:} Exp - Expert, Prof - Proficient, Com - Competent
        \item \textbf{\cmark:} Steps are applied correctly and the test case passed.        \textbf{\xmark:} Steps are either missed or applied incorrectly and the test case failed.      \textbf{--:} Steps not required for the dataset.																	
	\end{tablenotes}
     \vspace{-0.2cm}
\end{table*}

\begin{table*}[]
\huge
\caption{Summary of issues detected by {\em KUnit} using mock data in model design stage.}
\renewcommand{\arraystretch}{1}

\begin{adjustbox}{width=\textwidth,center}

\begin{tabular}{|cll|ccc|ccc|ccc|ccc|ccc|}
\hline
\rowcolor[HTML]{BFBFBF} 
\multicolumn{1}{|c|}{\cellcolor[HTML]{BFBFBF}}                                                                            & \multicolumn{2}{c|}{\cellcolor[HTML]{BFBFBF}}                                                                                                                                                                                                & \multicolumn{3}{c|}{\cellcolor[HTML]{BFBFBF}\textbf{Task 1 (Portfolio)}}                                                                                                                                                                                                              & \multicolumn{3}{c|}{\cellcolor[HTML]{BFBFBF}\textbf{Task 2 (Grain)}}                                                                                                                                                                                                                  & \multicolumn{3}{c|}{\cellcolor[HTML]{BFBFBF}\textbf{Task 3 (Truck)}}                                                                                                                                                                                                                   & \multicolumn{3}{c|}{\cellcolor[HTML]{BFBFBF}\textbf{Task 4 (Loan)}}                                                                                                                                                                                                                   & \multicolumn{3}{c|}{\cellcolor[HTML]{BFBFBF}\textbf{Task 5 (Train)}}                                                                                                                                                                                                                  \\ \cline{4-18} 
\rowcolor[HTML]{BFBFBF} 
\multicolumn{1}{|c|}{\multirow{-2}{*}{\cellcolor[HTML]{BFBFBF}\textbf{Stage}}}                                            & \multicolumn{2}{c|}{\multirow{-2}{*}{\cellcolor[HTML]{BFBFBF}\textbf{Tests}}}                                                                                                                                                                & \multicolumn{1}{c|}{\cellcolor[HTML]{BFBFBF}\textbf{\begin{tabular}[c]{@{}c@{}}S1\\  (Com)\end{tabular}}} & \multicolumn{1}{c|}{\cellcolor[HTML]{BFBFBF}\textbf{\begin{tabular}[c]{@{}c@{}}S2\\  (Prof)\end{tabular}}} & \textbf{\begin{tabular}[c]{@{}c@{}}S3\\  (Exp)\end{tabular}} & \multicolumn{1}{c|}{\cellcolor[HTML]{BFBFBF}\textbf{\begin{tabular}[c]{@{}c@{}}S1\\  (Com)\end{tabular}}} & \multicolumn{1}{c|}{\cellcolor[HTML]{BFBFBF}\textbf{\begin{tabular}[c]{@{}c@{}}S2\\  (Prof)\end{tabular}}} & \textbf{\begin{tabular}[c]{@{}c@{}}S3\\  (Exp)\end{tabular}} & \multicolumn{1}{c|}{\cellcolor[HTML]{BFBFBF}\textbf{\begin{tabular}[c]{@{}c@{}}S1\\  (Prof)\end{tabular}}} & \multicolumn{1}{c|}{\cellcolor[HTML]{BFBFBF}\textbf{\begin{tabular}[c]{@{}c@{}}S2\\  (Prof)\end{tabular}}} & \textbf{\begin{tabular}[c]{@{}c@{}}S3\\  (Exp)\end{tabular}} & \multicolumn{1}{c|}{\cellcolor[HTML]{BFBFBF}\textbf{\begin{tabular}[c]{@{}c@{}}S1\\  (Prof)\end{tabular}}} & \multicolumn{1}{c|}{\cellcolor[HTML]{BFBFBF}\textbf{\begin{tabular}[c]{@{}c@{}}S2\\  (Exp)\end{tabular}}} & \textbf{\begin{tabular}[c]{@{}c@{}}S3\\  (Exp)\end{tabular}} & \multicolumn{1}{c|}{\cellcolor[HTML]{BFBFBF}\textbf{\begin{tabular}[c]{@{}c@{}}S1\\  (Prof)\end{tabular}}} & \multicolumn{1}{c|}{\cellcolor[HTML]{BFBFBF}\textbf{\begin{tabular}[c]{@{}c@{}}S2\\  (Exp)\end{tabular}}} & \textbf{\begin{tabular}[c]{@{}c@{}}S3\\  (Exp)\end{tabular}} \\ \hline
\multicolumn{1}{|c|}{}                                                                                                    & \multicolumn{1}{l|}{}                                                                                                          & Check input shape in input layer is correct                                                                 & \multicolumn{1}{c|}{\cmark}                                                                                    & \multicolumn{1}{c|}{\cellcolor[HTML]{DAF2D0}\xmark}                                                             & \cmark                                                            & \multicolumn{1}{c|}{\cmark}                                                                                    & \multicolumn{1}{c|}{\cmark}                                                                                     & \cmark                                                            & \multicolumn{1}{c|}{\cellcolor[HTML]{DAF2D0}\xmark}                                                             & \multicolumn{1}{c|}{\cmark}                                                                                     & \cmark                                                            & \multicolumn{1}{c|}{\cmark}                                                                                     & \multicolumn{1}{c|}{\cellcolor[HTML]{DAF2D0}\xmark}                                                            & \cmark                                                            & \multicolumn{1}{c|}{\cmark}                                                                                     & \multicolumn{1}{c|}{\cmark}                                                                                    & \cmark                                                            \\ \cline{3-18} 
\multicolumn{1}{|c|}{}                                                                                                    & \multicolumn{1}{l|}{\multirow{-2}{*}{\begin{tabular}[c]{@{}l@{}}Data and model input-output \\  layer alignment\end{tabular}}} & Check output shape in input layer is correct                                                                & \multicolumn{1}{c|}{\cmark}                                                                                    & \multicolumn{1}{c|}{\cmark}                                                                                     & \cmark                                                            & \multicolumn{1}{c|}{\cellcolor[HTML]{DAF2D0}\xmark}                                                            & \multicolumn{1}{c|}{\cmark}                                                                                     & \cmark                                                            & \multicolumn{1}{c|}{\cmark}                                                                                     & \multicolumn{1}{c|}{\cmark}                                                                                     & \cmark                                                            & \multicolumn{1}{c|}{\cmark}                                                                                     & \multicolumn{1}{c|}{\cmark}                                                                                    & \cmark                                                            & \multicolumn{1}{c|}{\cmark}                                                                                     & \multicolumn{1}{c|}{\cmark}                                                                                    & \cmark                                                            \\ \cline{2-18} 
\multicolumn{1}{|c|}{}                                                                                                    & \multicolumn{1}{l|}{}                                                                                                          & \begin{tabular}[c]{@{}l@{}}Activation functions are applied correctly \\  in all hidden layers\end{tabular} & \multicolumn{1}{c|}{\cellcolor[HTML]{DAF2D0}\xmark}                                                            & \multicolumn{1}{c|}{\cellcolor[HTML]{DAF2D0}\xmark}                                                             & \cellcolor[HTML]{DAF2D0}\xmark                                    & \multicolumn{1}{c|}{\cmark}                                                                                    & \multicolumn{1}{c|}{\cmark}                                                                                     & \cmark                                                            & \multicolumn{1}{c|}{\cellcolor[HTML]{DAF2D0}\xmark}                                                             & \multicolumn{1}{c|}{\cellcolor[HTML]{DAF2D0}\xmark}                                                             & \cellcolor[HTML]{DAF2D0}\xmark                                    & \multicolumn{1}{c|}{\cmark}                                                                                     & \multicolumn{1}{c|}{\cellcolor[HTML]{DAF2D0}\xmark}                                                                                   & \cellcolor[HTML]{DAF2D0}\xmark                                    & \multicolumn{1}{c|}{\cmark}                                                                                     & \multicolumn{1}{c|}{\cmark}                                                                                    & \cmark                                                            \\ \cline{3-18} 
\multicolumn{1}{|c|}{}                                                                                                    & \multicolumn{1}{l|}{}                                                                                                          & \begin{tabular}[c]{@{}l@{}}Output layer format is correct \\  depending on the task\end{tabular}            & \multicolumn{1}{c|}{\cmark}                                                                                    & \multicolumn{1}{c|}{\cmark}                                                                                     & \cmark                                                            & \multicolumn{1}{c|}{\cmark}                                                                                    & \multicolumn{1}{c|}{\cmark}                                                                                     & \cmark                                                            & \multicolumn{1}{c|}{\cellcolor[HTML]{DAF2D0}\xmark}                                                             & \multicolumn{1}{c|}{\cmark}                                                                                     & \cmark                                                            & \multicolumn{1}{c|}{\cmark}                                                                                     & \multicolumn{1}{c|}{\cellcolor[HTML]{DAF2D0}\xmark}                                                                                   & \cellcolor[HTML]{FBE2D5}\xmark                                    & \multicolumn{1}{c|}{\cellcolor[HTML]{DAF2D0}\xmark}                                                             & \multicolumn{1}{c|}{\cmark}                                                                                    & \cmark                                                            \\ \cline{3-18} 
\multicolumn{1}{|c|}{}                                                                                                    & \multicolumn{1}{l|}{}                                                                                                          & Correct loss function is selected                                                                           & \multicolumn{1}{c|}{\cellcolor[HTML]{DAF2D0}\xmark}                                                            & \multicolumn{1}{c|}{\cmark}                                                                                     & \cmark                                                            & \multicolumn{1}{c|}{\cmark}                                                                                    & \multicolumn{1}{c|}{\cmark}                                                                                     & \cmark                                                            & \multicolumn{1}{c|}{\cmark}                                                                                     & \multicolumn{1}{c|}{\cmark}                                                                                     & \cmark                                                            & \multicolumn{1}{c|}{\cmark}                                                                                     & \multicolumn{1}{c|}{\cmark}                                                                                    & \cmark                                                            & \multicolumn{1}{c|}{\cmark}                                                                                     & \multicolumn{1}{c|}{\cmark}                                                                                    & \cmark                                                            \\ \cline{3-18} 
\multicolumn{1}{|c|}{}                                                                                                    & \multicolumn{1}{l|}{\multirow{-4}{*}{Correct operations}}                                                                      & Correct metrics is selected                                                                                 & \multicolumn{1}{c|}{\cellcolor[HTML]{DAF2D0}\xmark}                                                            & \multicolumn{1}{c|}{\cmark}                                                                                     & \cmark                                                            & \multicolumn{1}{c|}{\cellcolor[HTML]{DAF2D0}\xmark}                                                            & \multicolumn{1}{c|}{\cellcolor[HTML]{DAF2D0}\xmark}                                                             & \cmark                                                            & \multicolumn{1}{c|}{\cellcolor[HTML]{DAF2D0}\xmark}                                                             & \multicolumn{1}{c|}{\cellcolor[HTML]{DAF2D0}\xmark}                                                             & \cmark                                                            & \multicolumn{1}{c|}{\cellcolor[HTML]{DAF2D0}\xmark}                                                             & \multicolumn{1}{c|}{\cellcolor[HTML]{DAF2D0}\xmark}                                                            & \cellcolor[HTML]{DAF2D0}\xmark                                    & \multicolumn{1}{c|}{\cmark}                                                                                     & \multicolumn{1}{c|}{\cmark}                                                                                    & \cmark                                                            \\ \cline{2-18} 
\multicolumn{1}{|c|}{}                                                                                                    & \multicolumn{1}{l|}{}                                                                                                          & Model is learning and accuracy is changing                                                                  & \multicolumn{1}{c|}{\cmark}                                                                                    & \multicolumn{1}{c|}{\cmark}                                                                                     & \cmark                                                            & \multicolumn{1}{c|}{\cmark}                                                                                    & \multicolumn{1}{c|}{\cmark}                                                                                     & \cmark                                                            & \multicolumn{1}{c|}{\cmark}                                                                                     & \multicolumn{1}{c|}{\cmark}                                                                                     & \cmark                                                            & \multicolumn{1}{c|}{\cmark}                                                                                     & \multicolumn{1}{c|}{\cmark}                                                                                    & \cellcolor[HTML]{FBE2D5}\xmark                                    & \multicolumn{1}{c|}{\cmark}                                                                                     & \multicolumn{1}{c|}{\cmark}                                                                                    & \cmark                                                            \\ \cline{3-18} 
\multicolumn{1}{|c|}{}                                                                                                    & \multicolumn{1}{l|}{}                                                                                                          & No oscillating loss                                                                                         & \multicolumn{1}{c|}{\cmark}                                                                                    & \multicolumn{1}{c|}{\cmark}                                                                                     & \cmark                                                            & \multicolumn{1}{c|}{\cellcolor[HTML]{DAF2D0}\xmark}                                                            & \multicolumn{1}{c|}{\cellcolor[HTML]{DAF2D0}\xmark}                                                             & \cellcolor[HTML]{DAF2D0}\xmark                                    & \multicolumn{1}{c|}{\cmark}                                                                                     & \multicolumn{1}{c|}{\cmark}                                                                                     & \cmark                                                            & \multicolumn{1}{c|}{\cmark}                                                                                     & \multicolumn{1}{c|}{\cellcolor[HTML]{DAF2D0}\xmark}                                                                                   & \cellcolor[HTML]{FBE2D5}\xmark                                    & \multicolumn{1}{c|}{\cellcolor[HTML]{DAF2D0}\xmark}                                                             & \multicolumn{1}{c|}{\cellcolor[HTML]{DAF2D0}\xmark}                                                                                 & \cellcolor[HTML]{DAF2D0}\xmark                                    \\ \cline{3-18} 
\multicolumn{1}{|c|}{\multirow{-9}{*}{\begin{tabular}[c]{@{}c@{}}Model Design\\  (Testing using \\ \textbf{Mock Data})\end{tabular}}} & \multicolumn{1}{l|}{\multirow{-3}{*}{\begin{tabular}[c]{@{}l@{}}Performance of model \\  on mock data\end{tabular}}}           & Model is not converging slowly                                                                              & \multicolumn{1}{c|}{\cmark}                                                                                    & \multicolumn{1}{c|}{\cmark}                                                                                     & \cmark                                                            & \multicolumn{1}{c|}{\cellcolor[HTML]{DAF2D0}\xmark}                                                            & \multicolumn{1}{c|}{\cellcolor[HTML]{DAF2D0}\xmark}                                                             & \cellcolor[HTML]{DAF2D0}\xmark                                    & \multicolumn{1}{c|}{\cmark}                                                                                     & \multicolumn{1}{c|}{\cmark}                                                                                     & \cmark                                                            & \multicolumn{1}{c|}{\cellcolor[HTML]{DAF2D0}\xmark}                                                             & \multicolumn{1}{c|}{\cellcolor[HTML]{DAF2D0}\xmark}                                                                                   & \cmark                                                            & \multicolumn{1}{c|}{\cellcolor[HTML]{DAF2D0}\xmark}                                                             & \multicolumn{1}{c|}{\cellcolor[HTML]{DAF2D0}\xmark}                                                                                    & \cellcolor[HTML]{FBE2D5}\xmark                                    \\ \hline
\multicolumn{3}{|c|}{\textbf{Number of issues detected in model structure}}                                                                                                                                                                                                                                                                                              & \multicolumn{1}{c|}{\textbf{3}}                                                                           & \multicolumn{1}{c|}{\textbf{2}}                                                                            & \textbf{1}                                                   & \multicolumn{1}{c|}{\textbf{4}}                                                                           & \multicolumn{1}{c|}{\textbf{3}}                                                                            & \textbf{2}                                                   & \multicolumn{1}{c|}{\textbf{4}}                                                                            & \multicolumn{1}{c|}{\textbf{2}}                                                                            & \textbf{1}                                                   & \multicolumn{1}{c|}{\textbf{2}}                                                                            & \multicolumn{1}{c|}{\textbf{6}}                                                                           & \textbf{5}                                                   & \multicolumn{1}{c|}{\textbf{3}}                                                                            & \multicolumn{1}{c|}{\textbf{2}}                                                                           & \textbf{2}                                                   \\ \hline
\multicolumn{3}{|c|}{\textbf{Number of issues resolved by participants}}                                                                                                                                                                                                                                                                                                 & \multicolumn{1}{c|}{\textbf{3}}                                                                           & \multicolumn{1}{c|}{\textbf{2}}                                                                            & \textbf{1}                                                   & \multicolumn{1}{c|}{\textbf{4}}                                                                           & \multicolumn{1}{c|}{\textbf{3}}                                                                            & \textbf{2}                                                   & \multicolumn{1}{c|}{\textbf{4}}                                                                            & \multicolumn{1}{c|}{\textbf{2}}                                                                            & \textbf{1}                                                   & \multicolumn{1}{c|}{\textbf{2}}                                                                            & \multicolumn{1}{c|}{\textbf{6}}                                                                           & \textbf{2}                                                   & \multicolumn{1}{c|}{\textbf{3}}                                                                            & \multicolumn{1}{c|}{\textbf{2}}                                                                           & \textbf{1}                                                   \\ \hline
\multicolumn{3}{|c|}{\textbf{Number of issues dismissed as false alarms by participants}}                                                                                                                                                                                                                                                                                & \multicolumn{1}{c|}{\textbf{0}}                                                                           & \multicolumn{1}{c|}{\textbf{0}}                                                                            & \textbf{0}                                                   & \multicolumn{1}{c|}{\textbf{0}}                                                                           & \multicolumn{1}{c|}{\textbf{0}}                                                                            & \textbf{0}                                                   & \multicolumn{1}{c|}{\textbf{0}}                                                                            & \multicolumn{1}{c|}{\textbf{0}}                                                                            & \textbf{0}                                                   & \multicolumn{1}{c|}{\textbf{0}}                                                                            & \multicolumn{1}{c|}{\textbf{0}}                                                                           & \textbf{3}                                                   & \multicolumn{1}{c|}{\textbf{0}}                                                                            & \multicolumn{1}{c|}{\textbf{0}}                                                                           & \textbf{1}                                                   \\ \hline
\end{tabular}
\label{rq1userstudymodel}
\end{adjustbox}

\begin{tablenotes}
 \centering
        \fontsize{7.5pt}{7.5pt}\selectfont
        \item \textbf{Expertise level of participants:} Exp - Expert, Prof - Proficient, Com - Competent
        \item \textbf{\cmark:} Steps are applied correctly and the test case passed.        \textbf{\xmark:} Steps are either missed or applied incorrectly and the test case failed. 													
	\end{tablenotes}
 \vspace{-0.5cm}
\end{table*}
\subsubsection{Participants}
We recruited participants via LinkedIn using direct messages and 
university mailing lists that described our study and a link to the screener survey. We screened participants (i) who were over 18, (ii) who had at least one year of programming experience, and (iii) who had experience with DL programming. 
To evaluate the applicability of our approach in everyday scenarios and industry settings, we recruited both graduate students and 
industry professionals.
In total, we recruited 36 participants (21 male and 15 female), 24 graduate students from different universities, and 12 industry professionals working in Google, IBM, Neural Lab, \etc.
Participants were asked to self-classify their level of expertise on a scale from 1 (beginner)
to 5 (expert). The obtained expertise levels are: using existing DL programs ($\mu = 4.1, \sigma = 0.7$), developing new DL programs ($\mu = 4.0, \sigma = 0.8$), debugging DL programs ($\mu = 3.9, \sigma = 0.8$), and familiarity with preprocessing steps ($\mu = 4.0, \sigma = 0.8$).

In the study design, we aimed to have each task performed by participants with varying levels of expertise.
This allowed us to investigate the mistakes made by developers across different skill levels and evaluate the utility of unit testing in DL applications.
Participants were assigned tasks based on their self-reported expertise level. 
During self-assessment, we found that the participants rated themselves as either 3 (competent), 4 (proficient), or 5 (expert).
When a participant scheduled a session, we assigned a task that no other participant with the same experience level had already been assigned.  If a new participant with the same expertise level scheduled a time slot
and all tasks had been completed by others with the same
expertise level, we randomly assigned tasks to ensure that 
at least three different participants performed each task.

\subsection{Results}
In this section, we 
report on the efficiency of
our technique 
and answer our research questions.

\subsubsection{Do mock objects aid in testing each functionality in isolation without reliance on external dependencies?}
In this research question, we validate whether different functionalities of the DL programs can be tested without committing to a labeled dataset or model using mock objects, isolating the code being tested from external dependencies (dataset or model).
To answer this research question, we first evaluated the performance of {\em KUnit} on 50 DL programs in our benchmark. 
The first author
manually inspected each DL program and divided it into two parts: \textit{Data Preparation:} contains all the steps related to data preparation and \textit{Model Design:} contains all the steps related to designing the model including the compilation step.
Each part is tested independently using the mocks automatically generated by {\em KUnit}.
To address the challenges posed by the indeterministic nature of DL applications, we execute each test 3 times and consider issues reported in more than one run.
The buggy version of the original program was examined, and the number of issues in each of the 50 programs was counted. 
Table~\ref{SOFsummary} reports the total number of issues found in 50 programs across different categories.
The results shown in Table~\ref{SOFsummary} depict that the mock model facilitated testing of the data preparation steps, with {\em KUnit} identifying 10 out of 12 issues in this stage.
Similarly, for the model design stage, mock data facilitated testing of the designed model, with {\em KUnit} identifying 53 out of 62 issues in this stage.
Further investigation was done to determine the reason behind the issues missed by {\em KUnit}. 
In the data preparation stage, we found that the 2 missed issues were related to scaling the labels to the appropriate range to match the output layer activation. {\em KUnit} only verifies the scaling of the data, not the labels.
Therefore, {\em KUnit} missed these issues.
For the model design stage, the assertions used in {\em KUnit} are 
obtained from various sources (discussed in Section~\ref{bugconditions}) which
cover various frequently occurring scenarios that might not account for some edge cases, such as models designed for datasets with specific label ranges; {\em KUnit} missed 9 issues.
As {\em KUnit} is open-source, developers have the flexibility to refine these assertions or define new ones according to their needs.

Secondly, during the user study, participants used {\em KUnit} to independently test the data preparation steps and designed model using mocks. Tables~\ref{rq1userstudydata} and \ref{rq1userstudymodel} present the results of testing the three solutions (S1, S2, and S3) provided by different participants for each task, with columns indicating the issues detected in each solution.
For the data preparation stage, Table~\ref{rq1userstudydata} demonstrates that a total of 25 issues were identified by {\em KUnit} using mock models for all the tasks.
All these issues were acknowledged as valid findings
by participants with varying levels of expertise and were subsequently resolved by them.
For the model design stage, the participants tested the designed model using the mock data. 
In Table~\ref{rq1userstudymodel}, 
for each task, we highlighted the issues identified and reported by {\em KUnit}. 
Across all tasks, {\em KUnit} detected a total of 42 issues using mock data, of which 38 were accepted and resolved by participants with varying levels of expertise.
However, for 4 issues reported by {\em KUnit}, 2 participants mentioned that, based on their experience, these might be false alarms.
For example, in Task 4, a binary classification task, {\em KUnit} verifies that the output layer yields a value between 0 and 1 depicting the positive class probability (property of the sigmoid function).
However, for Task 4 (S3), the participant expressed a preference for using a probability distribution (softmax function) to represent the output layer results instead of class probabilities (sigmoid function), leaving room for extending the problem to multiclass classification in the future.  
Similarly, for detecting the oscillating loss issue, {\em KUnit} monitors the loss after every 5 epochs. However, for Task 5 (S3), the participant stated that they prefer to evaluate the model's stability every 10 epochs.
Therefore, due to differing criteria and preferences used by different developers for evaluating model stability, 4 out of 42 issues were dismissed as a false alarm by 2 participants for Task 4 (S3) and Task 5 (S3). Given that {\em KUnit} is open-source, developers can customize these assertions to fit their problem requirements.
\textit{In summary, our results demonstrate that mock objects facilitated independent testing of each functionality and assisted in the early detection of issues.}

\subsubsection{How efficient are mock objects in identifying issues compared to traditional deep learning testing approaches?}
In this research question, we evaluate the efficiency of mock objects in uncovering issues in DL programs at an early stage, that in current practice, are detected after combining different stages, specifically during training.
To compare the efficiency of {\em KUnit} with traditional DL testing approaches, we evaluated 50 programs in our benchmark.
The original program (with data and model stages combined) is tested using a state-of-the-art approach \cite{wardat22DeepDiagnosis} and the results are compared with the issues identified by {\em KUnit} at each stage.
DeepDiagnosis~\cite{wardat22DeepDiagnosis} is a fault localization tool that detects silent bugs in DL programs by monitoring for abnormal behavior during training.
This tool is selected because it covers most of the silent bugs (8) encountered during model training compared to other existing fault localization tools~\cite{schoop2021umlaut,wardat21DeepLocalize,Zhang21Autotrainer,cao2022deepfd}.
By comparing {\em KUnit} with DeepDiagnosis, we evaluate whether {\em KUnit} can effectively identify issues at an earlier stage using mocks before data-model integration that DeepDiagnosis identifies after integration during training.
Our analysis shows that, 
for 22 programs with issues in only one stage, \ie, data preparation or model design, {\em KUnit} identified the same problems with mocks that DeepDiagnosis identified using the original dataset during training. 
For the 25 programs with issues in both the data preparation and model design stages, DeepDiagnosis detected issues in the data preparation stage for 7 of these programs. 
However, uncovering bugs in the model design stage with DeepDiagnosis requires fixing data preparation issues first, necessitating multiple iterations to identify problems in the model design stage. 
Similarly, for 5 of these programs, DeepDiagnosis reported a numerical error in computation but could not determine the error-inducing stage. 
For the remaining 13 programs, DeepDiagnosis did not detect any issues.
The main challenge for DeepDiagnosis stems from programs with multiple bugs, as it detects one issue at a time.
This requires retraining the model on an original training dataset after every modification, leading to inefficient use of computational resources~\cite{Zhang21Autotrainer}.
In contrast, testing each stage separately with {\em KUnit} using mocks provides a lightweight emulation of dependencies, facilitating testing each stage before and after modifications, thereby saving resources.
As a result, {\em KUnit} detected issues in all 25 programs.
Some numerical computations in DNNs are highly data-dependent.
For example, using activation functions that are not suitable for certain input ranges can lead to out-of-range problems, which cannot be detected during unit testing with mocks.
Therefore, for 3 programs, {\em KUnit} missed these issues, whereas, testing using the original dataset helped DeepDiagnosis identify issues in 2 out of 3 programs. 

During the user study, participants utilized both {\em KUnit} and DeepDiagnosis for debugging. The results show that testing individual stages in isolation with mocks helped {\em KUnit} to efficiently pinpoint the root causes of issues.
In contrast, DeepDiagnosis, which detects issues after data-model integration, cannot identify the root cause of the issue in programs with multiple bugs. This is particularly evident in programs with issues originating from both the data preparation and model design stages, as these issues often exhibit overlapping symptoms during training.
Due to this, in 10 programs with multiple bugs, DeepDiagnosis reported a numerical error but failed to pinpoint the root cause. Detailed results are reported in the supplementary material~\cite{rq2results,rq2userstudyresults}.
We also analyzed the debugging time for each task in Tables~\ref{rq1userstudydata} and \ref{rq1userstudymodel}, comparing results with and without {\em KUnit}. In the traditional setting, when data and model are integrated and tested using DeepDiagnosis, we observed that the quality of the preprocessed data significantly impacted the debugging time. Participants had to resolve data-related issues before addressing model-specific problems. 
In contrast, when using {\em KUnit}, the data and model are tested in isolation using mocks, allowing participants to focus on stage-specific issues. As DeepDiagnosis identifies issues after integration, for a fair comparison, we computed the total time, \ie, summation of time taken by participants to resolve bugs in data and model stages using {\em KUnit} separately and compared it with DeepDignosis. 
Our analysis reveals that, on average, participants took 15 and 12 minutes to debug the DL programs using DeepDiagnosis, and {\em KUnit}, respectively. By isolating data and model, {\em KUnit} facilitates a more focused and efficient debugging process that can save time and resources during training. Detailed analysis is provided in our GitHub repository~\cite{debuggingtime}.
\textit{In summary, our analysis shows that mock objects effectively mimic essential system behaviors, simplifying complexity for unit testing and assisting in identifying issues that lead to abnormal behavior during training.
Mock testing is efficient and resource-friendly, especially for programs with issues in multiple stages.}

\subsubsection{How do developers perceive the effectiveness of unit testing using mocks compared to traditional deep learning testing approaches?}
\begin{figure}[t]
	\centering
	\includegraphics[scale=0.65]
 {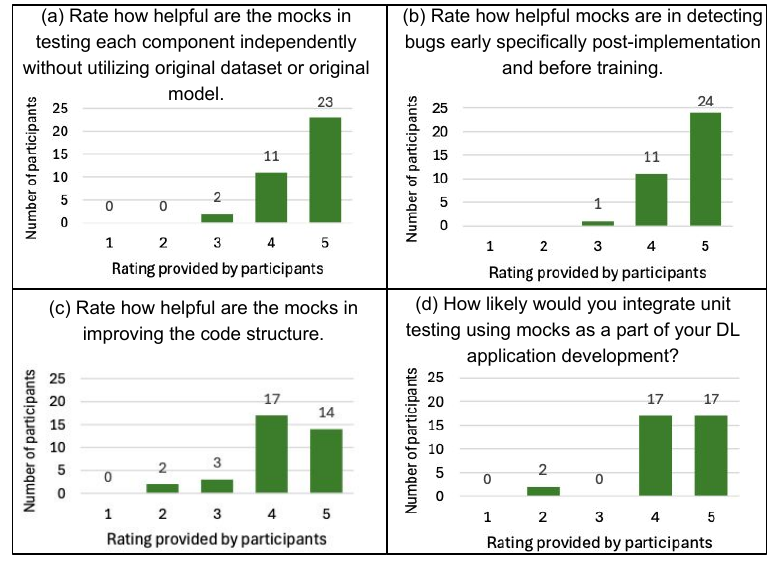}
    \caption{Survey results with participants ratings.}
	\label{Userratings}
\vspace{-0.4cm}
\end{figure}

To answer this research question, we collected survey responses from 36 participants during the user study.
In particular, on a 5-point Likert scale question, participants rated their experience about the usefulness of mocks for unit testing.
\figref{Userratings} shows the survey results with participants ratings.
As illustrated in \figref{Userratings}(a), 34 participants rated positively (rating $>$ 3) and found that mocks help test each component independently.
Likewise, 35 participants (\figref{Userratings}(b)) responded (rating $>$ 3)  that mocks facilitate the early identification and resolution of issues during the development process.
Regarding the usefulness of the mocks in improving code structure, 31 participants (\figref{Userratings}(c)) responded positively (rating $>$ 3), while 5 participants (rating $\leq$ 3) 
expressed the concern that mocks might produce false alarms and mislead efforts to improve the code structure.
Regarding the integration of the mocks into their DL development process, 34 participants (\figref{Userratings}(d)) rated positively (rating $>$ 3) and expressed their interest in incorporating unit testing with mocks to enhance software reliability.
We also asked participants to share open-ended feedback on the advantages and disadvantages of {\em KUnit}.
Two authors conducted an open coding phase over the qualitative responses~\cite{weiss1995learning} and grouped codes into different themes.
For advantages, our 
inductive thematic analysis identified 6 repeated themes in participants' qualitative responses.
The themes are: \textit{bugs can be detected early on, makes testing easy/easier to manage, time efficient/saves a lot of time, automation reduces human efforts, saving resources, great experience/helpful/useful}. 
For disadvantages, we found 2 repeated themes: 
\textit{implementation in the industry could be challenging/overhead to set up} and \textit{incorrect reports/false alarms.} 
Detailed qualitative responses are provided in the supplementary material~\cite{participantsresponse}. 

During the user study, we observed that the learning curve for {\em KUnit} depends on the user’s familiarity with DL concepts and experience with Keras. To help {\em KUnit} users easily understand the workflow, we provided detailed documentation and a running example in our GitHub repository~\cite{myRepo}. Participants in the post-study feedback illustrated that the well-structured documentation helped them easily understand {\em KUnit}’s functionality and workflow. After familiarizing themselves with the workflow, which on average takes 10-15 minutes, they only need to adjust the interfaces to align with their task. In the data preparation stage, the user loads the original dataset, applies preprocessing steps, and executes the test file. {\em KUnit} generates a mock model and feeds preprocessed data into it to verify data quality. In the model design stage, users build the DNN model and run the test. {\em KUnit} generates mock data and feeds it into the designed model to verify its correctness and compatibility with expected data properties. In the post-study feedback, participants mentioned that they found {\em KUnit} easy to use with minimal manual effort required for setup and customization. They were able to adjust the interfaces to align with their task and used the framework for developing and unit testing their DL application.
By making {\em KUnit} open-source, we provide flexibility to adapt the tool to developer needs and improve its accuracy and ease of use through community contributions.
\textit{In summary, we found that the developers view unit testing using mocks as a valuable addition to traditional DL testing techniques. It enables independent testing of each component, facilitates early problem detection during development, and contributes to improving the overall code structure.}

\section{Threats to Validity}
\label{sec:THREATSTOVALIDITY}
A potential threat to the internal validity of our study is the possibility of bugs in {\em KUnit}'s implementation, which could lead to inaccurate results. To mitigate this risk, we conducted a user study involving developers with diverse expertise levels and backgrounds, ensuring diverse perspectives on {\em KUnit}'s functionality. Additionally, we have made {\em KUnit}'s source code publicly available, allowing other researchers to review and validate our work.
In the user study, participants used DeepDiagnosis for debugging. DeepDiagnosis detects bugs and suggests fixes to resolve them. Participants manually addressed each bug identified by DeepDiagnosis and repeated the process until no further issues were reported. To mitigate the risk due to incorrect tool usage, we had our protocol reviewed by the authors of DeepDiagnosis to ensure it aligned with its intended functionality~\cite{commwith}.
Our proposed approach may be affected by external threats, such as imprecise conditions used as assertions and the effectiveness of the actionable fixes provided as solutions. To mitigate the risk, we have adopted guidelines from previous works~\cite{islam20repairing,wardat22DeepDiagnosis,Zhang21Autotrainer,ahmed23dlcontract,Khairunnesa2023,cao2022deepfd,ghanbari2023deepmufl} and Keras documentation~\cite{Keras,kerasexamples}. 
In our empirical evaluation, we assessed 50 DL programs in our benchmark. This process involved manually separating each DL program into two parts: data preparation and model design, which could introduce human error. To mitigate this threat, one of the co-authors thoroughly reviewed each part to 
ensure correctness.

\section[Related]{Related Work}

\label{sec:relatedwork}
\subsection{Unit Testing using Mocks}
Unit testing stands as a fundamental practice in software development,
aimed at evaluating each functionality of the software independently and uncovering bugs early in the development cycle.
Due to dependencies, testing the code in isolation becomes challenging.
To tackle this challenge, a technique known as mock objects have been proposed in the past by Mackinnon \etal~\cite{mackinnon2000endotesting} for unit testing, involving the replacement of dependencies with dummy implementations.
Their findings suggest that using mock objects for unit tests improves test quality and the structure of both domain and test code.
Prior studies have shown the benefits of using mock objects for unit testing various applications, including servlet~\cite{thomas2002mock}, multi-agent systems~\cite{coelho2006unit}, mobile apps~\cite{fazzini2020framework}, and database applications~\cite{taneja2010moda}.
However, the use of mock objects for testing DL applications has not been investigated before.

\subsection{Fault Localization and Bug Repair in DL Programs}
The rise in DL application usage has led researchers to adapt fault localization techniques to validate DL systems and identify faulty behaviors. In the past, various static and dynamic analysis approaches were proposed for DL programs.
\nlint~\cite{nikanjam2021neuralint} is a static analysis approach for automatic fault detection using predefined rules.
UMLAUT~\cite{schoop2021umlaut} combines static and dynamic analysis to examine program structure before training and model behavior during training.
DeepLocalize~\cite{wardat21DeepLocalize} is a dynamic fault localization approach that identifies 
numerical errors during training.
AutoTrainer~\cite{Zhang21Autotrainer} is a system designed to identify and repair 5 common training issues in DL models. 
DeepDiagnosis~\cite{wardat22DeepDiagnosis} is a dynamic 
technique that identifies various symptoms during training and suggests actionable fixes.
DeepFD~\cite{cao2022deepfd} is a learning-based framework for fault diagnosis.
TheDeepChecker \cite{BraiekDeepChecker} is a property-based debugging approach that detects bugs before, during, and after training.
deepmufl~\cite{ghanbari2023deepmufl} is a mutation-based fault localization approach that generates mutants of pre-trained models to detect bugs.
Prior works treat data preprocessing steps and the model as a comprehensive DL program, 
and identify and localize bugs by monitoring the training process.
In contrast, {\em KUnit} treats data and the model as independent entities and aims to detect bugs before integrating them.

\section[Conclusion]{Conclusion}
This paper introduces the concept of mock testing in the context of DNNs and presents a novel technique, {\em KUnit}. 
The technique is based on the idea of decoupling to reduce the dependencies between different stages of DL applications, specifically data preparation and model design. 
Decoupling is achieved by defining interfaces 
that facilitate the creation of mock objects for unit testing of each stage.
The empirical evaluation using 50 DL programs shows that in the data preparation stage, the mock model helped identify 10 issues, and mock data
assisted in identifying 53 issues in the model design stage.
In a user evaluation with 36 participants, {\em KUnit} helped resolve 25 issues in the data preparation stage and 38 issues in the model design stage.
Our results show that mock objects provided a lightweight emulation of the dependencies for unit testing and
identified issues that, in current practice, are identified after data and model integration, specifically during DL model training.
Participants found {\em KUnit} helpful for identifying and resolving issues early in the development process.

\label{sec:conclusion}

\section[Data Availability]{Data Availability}
\label{sec:data-availability}
Our evaluation results and the code for our framework, \textit{KUnit}, are available in our replication package \cite{myRepo}.
\section[Acknowledgment]{Acknowledgment}
\label{sec:Acknowledgment}

The authors thank the anonymous ICSE 2025 reviewers for their valuable feedback. This work is supported by the US National Science Foundation (NSF) under grants 2512857, 2512858, 15-18897, 15-13263, 21-20448, 19-34884, and 22-23812, the Fonds de Recherche du Quebec (FRQ), the Canadian Institute for Advanced Research (CIFAR), and the Natural Sciences and Engineering Research Council of Canada (NSERC).
All opinions are those of the authors and do not reflect the views of the sponsors.

\balance
\bibliographystyle{IEEEtran}
\bibliography{paper}

\end{document}